\documentclass[aps,prb,showpacs,raggedbottom,nobalancelastpage,amssymb,twocolumn,floatfix]{revtex4}
\usepackage{amsmath}
\usepackage{amsfonts}
\usepackage{graphicx}
\usepackage{float}
\usepackage{appendix}
\usepackage{dsfont}
\usepackage{amssymb}
\usepackage{bm}
\usepackage{color}
\usepackage{ulem}
\usepackage{soul}
\usepackage{natbib}
\setstcolor{red}
\usepackage{nccmath}
\usepackage{array}
\usepackage{placeins}

\newcommand{\beq}{\begin{equation}}
\newcommand{\eneq}{\end{equation}}
\newcommand{\be}{\begin{equation}}
\newcommand{\ee}{\end{equation}}
\newcommand{\bea}{\begin{eqnarray}}
\newcommand{\eea}{\end{eqnarray}}

\usepackage{multirow}
\usepackage{wasysym}

\begin{document}
\title{Lindblad equation approach to the determination of the optimal working point in nonequilibrium stationary states of an interacting electronic  one-dimensional
system: Application to the spinless Hubbard chain in the clean and in the weakly disordered limit}
\author{Andrea Nava, Marco Rossi,  and  Domenico Giuliano}
\affiliation{
Dipartimento di Fisica, Universit\`a della Calabria, Arcavacata di Rende I-87036, Cosenza, Italy \\
INFN - Gruppo collegato di Cosenza,
Arcavacata di Rende I-87036, Cosenza, Italy
 }
\date{\today}

\begin{abstract}
Using the Lindblad equation approach,  we derive the range of the parameters of an interacting one-dimensional electronic 
chain connected to two reservoirs in the large bias limit in which an optimal working point (corresponding to a change in the monotonicity of the stationary current as a function of the applied bias) emerges in the nonequilibrium stationary state. 
   
In the specific case of  the one-dimensional spinless fermionic Hubbard chain,  we prove that
an optimal working point emerges in  the dependence  of the stationary current
on the coupling between the chain and the reservoirs, both 
in the interacting and in the noninteracting case. 

We show that the  optimal working point is robust against 
localized defects of the chain, as well as against a limited 
amount of quenched disorder.

Eventually, we discuss the importance of our results for 
optimizing the performance of a quantum circuit by tuning
its components as  close as possible to their optimal working point.

\end{abstract}

\maketitle

\section{Introduction}
\label{intro}

Connecting a mesoscopic device with two or more electronic reservoirs, biased at different 
temperatures or chemical potentials, gives rise to finite currents flowing through the system, 
between the reservoirs.
At a low bias, the (equilibrium) transport  properties of the 
system are well accounted for, both analytically and numerically, 
within  linear response approach  \cite{mello}. Conversely, when 
the system is driven to the nonequilibrium regime (large bias), 
a fully analytical approach is
in practice unfeasible, due to the strong dependence of the 
dynamics   on the system and on the reservoirs separately, as well 
as on the nature of the coupling between the two of them \cite{prosen_1}. 

Despite the technical difficulties to  approach it, the
large bias  regime  is of great interest, as it is directly related to, e.g.,  
control of heat flow \cite{chf}, as well as quantum information processing \cite{qip}. 
For this reason, a large number of theoretical approaches has been developed to investigate the nature 
of nonequilibrium states, both in noninteracting,  as well as in interacting systems, such as 
the Landauer-Buttiker formalism \cite{lb,lb_2},  the quantum master 
equation approach \cite{qm,lind_1}, the renormalization group techniques \cite{rg_1,rg_2} (including the 
functional renormalization group approach \cite{addi.1}), and the bosonization methods \cite{boso_1,boso_2,boso_3}.
Yet, each of these methods has only a limited  range of applicability and, typically, none of them is able to 
fully catch all the relevant aspects of nonequilibrium physics, due to  the complexity of the systems, to the strength of the interaction,
to the peculiar nature of the  stationary states  that eventually set in, {\it et cetera}.

Therefore, to 
fully recover the nonequilibrium physics, one has to resort to 
a fully numerical approach, possibly complemented, when possible, 
with (approximate) analytical methods. In fact, when connected to 
one, or more, reservoirs,  ({\it{open}}) quantum systems can be described through a master
equation, aimed to represent the true quantum evolution after integrating over the reservoir degrees of freedom. 
The derivation of such equation usually relies on the so-called Markovian approximation, which consists of neglecting
memory effects under the assumption that the bath relaxation time is much shorter than the characteristic timescales
of the system \cite{petruccione,weiss}. The Lindblad equation (LE) \cite{lind_1} is among the most used master equations:
it stems from modeling the reservoirs as local ``jump'' operators injecting, or removing, 
 particles through the boundaries of the system. Aside of its universality, the LE can be readily 
approached  by means of a number of numerical methods, including  
  Quantum Monte Carlo techniques \cite{ben_1,ben_2,rossini}, time-dependent density matrix renormalization group ($t$-DMRG) \cite{vid_1,vid_2,feiguin,noneq} 
or current density functional approach \cite{rm_1}. 

Most of the literature concerning the numerical approach to the LE has been focused onto 
quantum $XXZ$ spin chains connected to reservoirs. The $XXZ$ spin-1/2 spin chain, indeed, 
provides a remarkable paradigmatic model where to address several key issues, such as 
the interplay between integrability breaking  and asymptotic evolution of the system towards 
an appropriate nonequilibrium stationary state (NESS) \cite{rossini_0}, the characterization of the NESS by looking 
at the real-space average magnetization in the $z$-direction and at the stationary spin current 
flowing through the chain,  the effects of isolated impurities as well as of quenched 
disorder on the NESS \cite{noneq,xxz_sabetta}, and so on. In addition, the quantum $XXZ$ spin chain is well-known 
to map onto a  model for spinless interacting one-dimensional lattice fermions (which throughout the 
paper we dub ``spinless fermionic Hubbard model'' (1HM), according to the widely used terminology \cite{spinl_1,spinl_2}), 
via Jordan-Wigner transformation \cite{jwigner}. Finally, it has been shown that it is 
possible to realize  quantum spin-1/2 $XXZ$ spin chains
with tunable impurities at pertinently engineered junctions of 
one-dimensional Josephson junction arrays \cite{giuso_1,giuso_2,giuso_3,giuso_4}.

A typical feature characterizing the NESS in the $XXZ$ spin-1/2 spin chain is 
the tendency of the system, driven out of equilibrium, to develop ferromagnetic 
domains, separated by domain walls that conspire to reduce the spin-flip rate and, 
therefore, to reduce the nonequilibrium stationary spin current through the chain \cite{rossini_0,xx_model}.
Similarly, the NESS in the 1HM is characterized by the emergence of 
domains in the chain with a uniform charge distributions, separated by 
charge domain walls that generate a counterfield reducing the total current
$I_{\rm st}$ flowing across the system in the NESS. The  
 domain walls give  rise, at large enough bias between the reservoirs, to a remarkable negative differential 
conductance (NDC), that is, to a region in which the current decreases if the bias increases.
The NDC is a feature of mesoscopic systems, such as semiconductor superlattices \cite{owp9}, 
interacting quantum gases \cite{owp8},  single molecule junctions \cite{owp10}, carbon nanotubes \cite{owp11}, 
 graphene transistors \cite{owp12}, and quantum dots coupled to electrodes \cite{sch}. 
Also, it can arise as  an effect of electron-phonon coupling 
\cite{zaz}. In  a quantum chain driven out of equilibrium the NDC emerges as a combined effect of 
the coherent many-body correlations and the incoherent charge pumping 
in the chain from the reservoirs \cite{rossini_0}.

 In general,  quantum transport properties in many-body systems strongly depend on the interplay 
 between bulk hopping processes, electron-electron interaction, noise and impurity 
 distribution, and boundary driving strength \cite{nazarov_blanter_2009,owp13,owp23}. 
 Typically, at small bias, the  current induced in the system exhibits Ohmic 
 law behavior, linearly increasing with increasing applied bias. The change in the monotonicity of 
$I_{\rm st}$ as a function of the applied bias necessarily implies the 
existence of an ''optimal working point`` (OWP), at which $I_{\rm st}$
takes the maximum value, given the other system parameters. 
In the 1HM the dependence of the OWP on the interaction has been 
extensively discussed in Ref.[\onlinecite{rossini}] (a similar analysis in 
the $XXZ$ chain has been performed in Ref.[\onlinecite{rossini_0}]).  It
has been found that, if the coupling strengths between the reservoirs 
and the 1HM are fixed, a necessary condition to recover the OWP is
having a nonzero interaction between the electrons. More generally, 
making a quantum system that is part of a quantum circuit work 
at the OWP, means maximizing the current flow supported by that 
part of the circuit at a given bias. Identifying the OWP for each 
component of the circuit is a necessary preliminary step to eventually 
make the circuit operate at its maximum possible efficiency. 
Moreover, an analogous optimization procedure for, e.g., the energy 
transport would have striking consequences for optimizing the 
control of the energy transfer between different part of mesoscopic
devices. In addition, the OWP can be associated with    nontrivial effects like the tendency 
 to enhance nonuniformities \cite{owp2,owp15,owp16} or the Gunn effect \cite{owp19,owp20,owp21}.
 Finally,  
it is worth noticing how the emergence of the OWP is 
 a typical behavior found in   fundamental traffic flow diagrams \cite{trafic_1,trafic_2},
 where the free flow phase and the congested phase are separated by an optimal value of 
the density, at which the traffic flow (the ``current'') is maximum. In fact, this observation
would suggest that a quantum chain (or a network) at large bias might potentially work as a ``quantum simulator''  of the 
fundamental traffic flow diagram, with potentially countless applications to 
real-life problems. 
Therefore, characterizing the OWP and its emergence as a function of the system parameters
is of the utmost importance for the implementations of controlled quantum circuits 
 \cite{owp1,owp3,owp17,owp22,heat_transport}. It is, therefore, crucial
to extend the analysis of Ref.[\onlinecite{rossini}] to a larger 
manifold in parameter space, which should possibly include 
parameters such as the  coupling strengths between the reservoirs 
and the 1HM, the amount of disorder in the system, and so on.

In this paper we systematically analyze the emergence and the characteristics of the OWP in the current 
$I_{\rm st}$ in the NESS  in interacting one-dimensional electronic systems connected 
to reservoirs, in the large bias limit. Complementing and extending the analysis of Ref.[\onlinecite{rossini}], 
we search for the OWP by considering how $I_{\rm st}$ changes as a function of 
the coupling strengths between the reservoirs  and the electronic system.  
To drive the system toward the NESS, we   implement the Lindblad master 
equation for a graph of $N$-sites connected with two, or more reservoirs.
In particular, we treat the interaction within  mean field (MF) approximation. 
Verifying, when possible,  the consistency of our results with the one 
already present in the literature, we check that, while allowing us for 
considerably simplifying  the calculations, 
our method   enables us to catch all the fundamental 
features characterizing the NESS, such as the 
dependence of $I_{\rm st}$ on the system parameters 
and the stationary distribution of the particle 
density in real space. 

Specifically, after presenting our approach in the general case of an interacting electronic 
system defined on a  graph connected to an arbitrary number of  reservoirs, 
we address the case study of a 1HM in the large bias limit, first with  homogeneous
system parameters,  then adding a single (``site''- or ``bond''-) 
impurity to the chain, and eventually in the presence of a finite amount of quenched disorder
in the system. In all 
the cases we focus on, we characterize the NESS in terms of the dependence of  $I_{\rm st}$ 
on the  coupling strengths between the reservoirs  and the electronic system,
and of the stationary distribution of the particle  density in real space. Doing so, we 
show that, as a function of the  tuning parameters,  the OWP emerges at the NESS
even in the absence of electronic interaction. Moreover, we directly check that, in the 
noninteracting, as well as in the interacting case, the OWP is pretty robust against
defects in the chain (isolated impurities), as well as against a moderate 
amount of quenched disorder. Independently tuning the interaction strength 
and the amount of disorder, we construct  the phase diagram of the system in the disorder - interaction
strength parameter space and, in particular, we draw the transition region beyond which 
the OWP disappears, $I_{\rm st}$ becomes zero and the whole system undergoes a 
Griffiths-like transition from a conducting to an insulating phase. Eventually, 
we check that our phase diagram is consistent with the one derived in 
Ref.[\onlinecite{noneq}]. 

Besides characterizing the NESS and the emergence of the OWP, taking advantage 
of the simplicity and of the effectiveness of our method, we can follow the evolution 
in time of our system toward the NESS, with no need for running long lasting 
numerical simulations. This allows us to map out, in various cases of interest, 
the details of how the 1HM evolves toward the NESS in real time. In particular, 
doing so we argue how the NESS is largely independent of the state  
we begin with and, in this respect, how it is a ``universal'' property of 
the chain-plus-reservoir system. Also,  we indirectly address 
 the interplay between the integrability and the evolution of the system 
toward the NESS. In general, the conservation laws associated to the integrability \cite{rossi_1,rossi_2,rossi_3}
are known to prevent the system from thermalizing toward a state characterized by a macroscopic hydrodynamical 
behavior in its transport properties \cite{prev_1,prev_2}. Breaking the integrability by adding a local 
impurity term to the otherwise integrable Hamiltonian should definitely trigger an evolution 
 toward a well defined NESS \cite{rigol}. Yet, we directly verify that the evolution  and the 
NESS itself barely depend on whether the chain is homogeneous, or with an isolated impurity. 
 This highlights how coupling the chain to the reservoirs already breaks 
the integrability, thus letting the system evolve toward the NESS and, therefore, how, in this 
specific case, adding an additional impurity to the chain has very little effect, if none at all, 
on the time evolution toward the NESS and on the NESS itself. 

The paper is organized as follows:
 
\begin{itemize}
 \item In Section \ref{lind} we present the model Hamiltonian for an interacting electronic system 
 defined over a generic graph. We therefore discuss our MF approach to the electronic 
 interaction, apply it to the graph system Hamiltonian and derive the corresponding LE. 
Eventually, we write down the conditions defining the NESS and explicitly solve 
them in the absence of interaction.
 
 \item In Section \ref{chain} we present a specific application of our approach to 
 a noninteracting electronic chain connected to two reservoirs at its endpoints. By sampling $I_{\rm st}$ and the charge density 
 distribution from the equilibrium to the large bias limit, as well as by varying the 
strength of the coupling between the chain and the reservoirs, we show that an OWP emerges, 
even in the absence of interaction, when considering $I_{\rm st}$, as a function of 
the coupling strength, taken in the large bias limit.

\item In Section \ref{interaction} we extend the analysis of Section \ref{chain} to the 
1HM connected to two reservoirs at the endpoints of the chain, at a generic value of 
the electronic interaction. Doing so, we evidence the rich set of phases generated in 
the system by turning on the interaction, by particularly focusing on the conductor to insulator
phase transition that emerges, in the large bias limit, at strong enough values of the 
interaction itself, and on its effects on the OWP.

\item In Section \ref{singleimp}   we analyze how adding an impurity 
term to the homogeneous 1HM Hamiltonian   affects, in the large bias limit,  the evolution of the system toward the NESS and 
the NESS itself. Specifically, we focus onto two different types of impurities: a ``site'' impurity, realized by 
 altering on a single site the otherwise uniform chemical potential, and a ``bond'' impurity, 
realized by changing the electronic hopping strength of a single bond of the chain.  

 \item In Section \ref{disorder} we analyze the effects of a finite density of impurities (quenched  disorder) in the 1HM 
by systematically discussing the phase diagram of the system,
in the large bias limit, in the disorder strength - interaction space, and how the NESS and the OWP 
are affected by the simultaneous presence of a finite disorder strength and of a nonzero 
electronic interaction.

\item In Section \ref{conc} we summarize our results and discuss possible further perspectives of our work.

\item In Appendix \ref{varia} we present a simple variational calculation that, despite its simplicity, 
is able to qualitatively 
catch the main features of the NESS that emerges in the chain in the large bias limit, both in 
the homogeneous case and in presence of a single site impurity.
 
\end{itemize}

\section{Model Hamiltonian and Lindblad equation}
\label{lind}

\begin{figure}
\center
\includegraphics*[width=0.8 \linewidth]{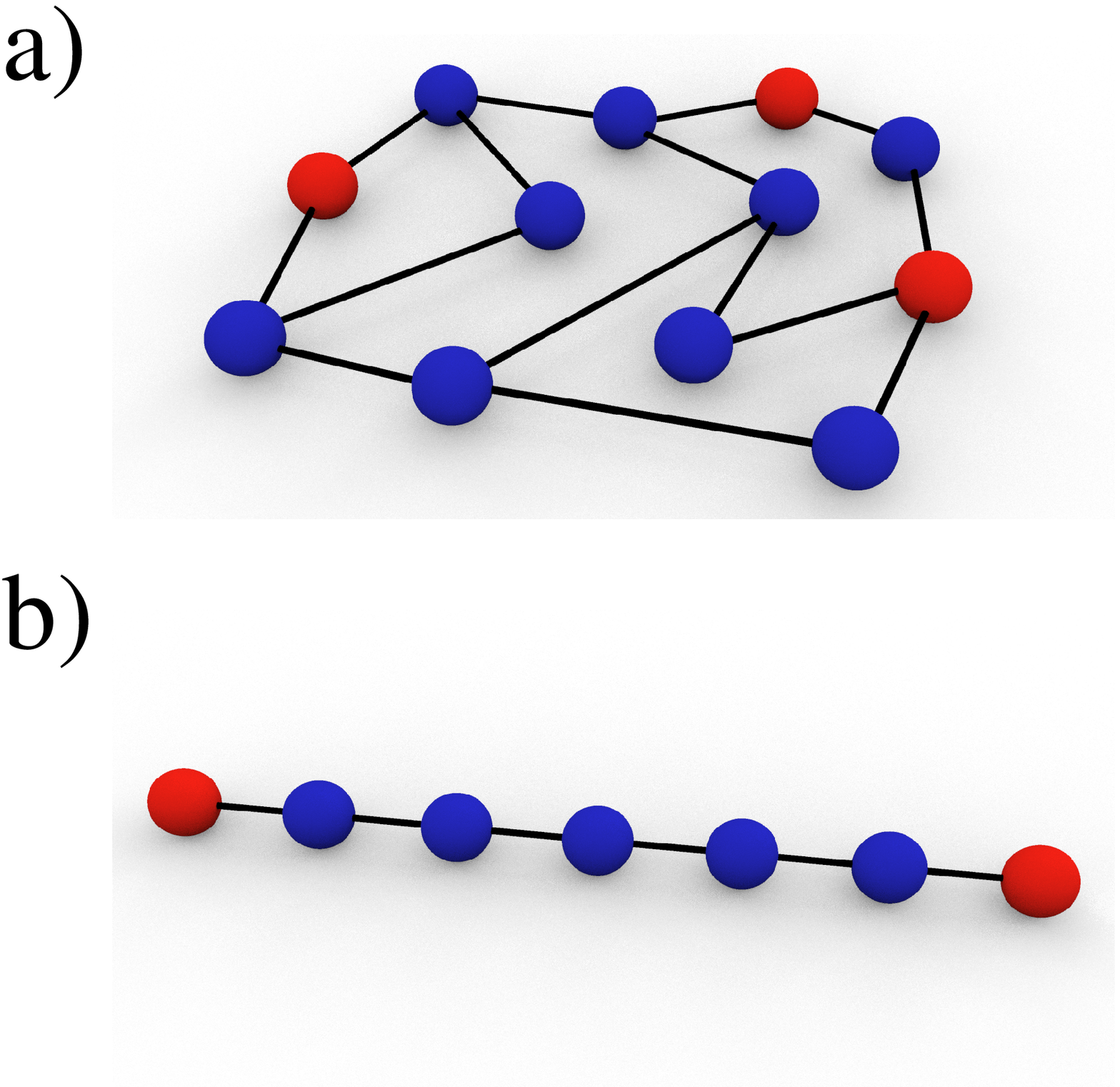}
\caption{\\ {\bf a)}: Sketch of a generic graph described by the Hamiltonian $H$ in Eq.(\ref{eq:deltastar-H}): 
the blue dots represent  generic sites connected to ``internal'' sites only, while the red dots represent sites connected to 
the external reservoirs, as well. A straight line connecting two sites 
represents a nonzero hopping strength and/or a nonzero interaction strength between the two sites; \\
{\bf b)}: The special graph we discuss in detail in our paper:  a linear chain connected to two reservoirs
at its endpoints through two red dots.}
\label{device}
\end{figure}

As a model Hamiltonian for a system of spinless, interacting electrons over an $N$-site 
graph, we use $H$, given by
 
\beq
H   =   -\sum_{j \neq k=1}^{N}J_{j,k}c_{j}^{\dagger}c_{k}-\sum_{j=0}^{N}\mu_{j}c_{j}^{\dagger}c_{j} +\sum_{j \neq k=1}^{N}U_{j,k}n_{j}n_{k}
\:\:\:\: . 
\label{eq:deltastar-H}
\eneq
\noindent
In Eq.(\ref{eq:deltastar-H}), $c_j , c_j^\dagger $ are respectively the single-fermion creation and annihilation operator 
at lattice site $j$, satisfying  the canonical anticommutation relations $\left\{ c_{j},c_{j'}^{\dagger}\right\} =\delta_{j,j'}$.
$n_j = c_j^\dagger c_j$ is the fermion number operator at site $j$. 
$J_{j,k}$ is the single-fermion hopping strength between sites $j$ and $k$, $\mu_j$ is the chemical potential at 
site $j$, $U_{j,k}$ is the density-density interaction strength between sites $j$ and $k$.   Eq.(\ref{eq:deltastar-H}) comprises the most 
general spinless Hubbard-like Hamiltonian for lattice spinless fermions   (see, e.g., Ref.[\onlinecite{essler}] for a comprehensive review 
about the one-dimensional Hubbard model). In principle, we allow any site of the lattice to be connected to an 
external reservoir. In Fig.\ref{device}{\bf a)},  we provide a sketch of 
the corresponding graph, with the blue dots representing generic sites and the red dots sites connected to 
the external reservoirs, as well as to other sites of the graph. A straight line connecting two sites 
represents a nonzero hopping strength and/or a nonzero interaction strength between the two sites.
Following the same drawing code, in Fig.\ref{device}{\bf b)} we show the simple graph representing 
the system on which  we focus most of the discussion of the following Sections: a linear chain connected to two reservoirs
at its endpoints.

\noindent
To describe the dynamics of the system represented by $H$ in Eq.(\ref{eq:deltastar-H}), once it is connected 
 to the external reservoirs,  we resort to the master equation (ME) approach to open quantum systems \cite{lind_1,lind_2}. 
Within the ME equaton framework, we derive the effective dynamics of the system by integrating 
over the reservoir degrees of freedom. Doing so, we resort to the   so-called Markovian approximation,  consisting in 
 neglecting memory effects under the assumption that the
bath relaxation time is much shorter than the characteristic
timescales of the system \cite{petruccione,weiss}.  Eventually, we derive the LE
for the open system  connected to the reservoirs. The LE is among the
most used master equations \cite{lind_3,lind_4,lind_5,moca}: its general form   consists of  a first order
differential equation for the time evolution of the system density matrix $\rho(t)$, 
given by 
 
\beq
\dot{\rho} ( t  )=
-i [H,\rho (t)  ]+\sum_{k} (L_{k}\rho ( t )  L_{k}^{\dagger}-\frac{1}{2} \{ L_{k}^{\dagger}L_{k},\rho (t)  \}  )
\:\:\:\: . 
\label{eq:lindbladeq}
\eneq
\noindent
The first term at the right-hand side of Eq.(\ref{eq:lindbladeq})  is the so-called
Liouvillian that describes the unitary evolution determined  by $H$. The second term, the so-called Lindbladian, includes
dissipation and decoherence on  the system dynamics. It 
depends on the so-called  ``jump'' operators $L_k$, which are determined by the
coupling between the system and the reservoirs.
Specifically, the  Liouvillian describes the unitary evolution brought by
$H$, while the Lindbladian includes the 
dissipation and the decoherence in the dynamics. 

In the following, we   consider  reservoirs that locally inject to,  or extract fermions from a generic site $j$ of 
the lattice, at  given and fixed rates. Consistently, we describe the injecting and extracting reservoirs at site $j$ in 
terms of the   Lindblad operators $L_{in,j} $ and $L_{out,j} $, given by 

\begin{eqnarray}
L_{in,j} & = & \sqrt{\Gamma_{j}}c_{j}^{\dagger} \nonumber \\
L_{out,j} & = & \sqrt{\gamma_{j}}c_{j} \:\:\:\: ,
\label{eq:deltastar-L}
\end{eqnarray}
\noindent
with $\Gamma_j$  and $\gamma_j$  being  the coupling strengths respectively determining 
the creation and the    annihilation   of a fermion at  site $j$.

Once we  determine  $\rho ( t )$ by solving Eq.(\ref{eq:lindbladeq}), we compute 
the (time dependent) expectation value of any observable $O$, $ O  ( t )  $ 
using

\beq
  O  ( t )  =
{\rm Tr} [ O \rho\left(t\right) ] 
\:\:\:\: . 
\label{ave.1}
\eneq
\noindent
Taking into account Eq.(\ref{ave.1}) and using the identities

\begin{eqnarray}
\left[A,BC\right] & = & B\left[A,C\right]+\left[A,B\right]C\nonumber \\
\left[A,BC\right] & = & \left\{ A,B\right\} C-B\left\{ A,C\right\} \nonumber  \\
{\rm Tr}\left(\left[A,B\right]\right) & = & 0
\;\;\;\; , 
\label{com.1}
\end{eqnarray}
\noindent
with $A , B , C$ being operators acting over the same Hilbert space, 
we  employ Eq.(\ref{eq:lindbladeq}) to write down the ME directly 
for $O ( t ) $. Specifically, we obtain

\begin{eqnarray}
\frac{d}{dt} O ( t )   & = & {\rm Tr}
 ( O  \dot{\rho} (t )  )= i {\rm Tr} [[  H, O   ] \rho ( t )  ] 
\label{eq:observable-master equation}  \\ 
& + & \sum_{k} ( {\rm Tr} [  L_{k}^{\dagger} O  L_{k}\rho ( t ) ]  -
\frac{1}{2} {\rm Tr} [  \{ L_{k}^{\dagger}L_{k}, O 
 \} \rho ( t ) ]      )\nonumber
\:\:\:\: . 
\end{eqnarray}
\noindent
For the sake of our analysis, in the following we will need to compute the average 
value of the occupation number for a generic site $j$ of the system, $n_j ( t ) = 
{\rm Tr} [ n_j  \rho ( t )]$, as well as 
of the currents flowing from the reservoirs into the site $j$,  $I_{in,j} ( t ) $,  or from site $j$ to 
the reservoir, $I_{out,j} ( t )$.  These are given by 
 
\begin{eqnarray}
I_{in,j} ( t )  & = & \Gamma_{i} (1-  n_j ( t )) \nonumber \\
I_{out,j}( t )  & = & \gamma_{i}  n_j ( t ) 
\:\:\:\: , 
\label{curave.1}
\end{eqnarray}
\noindent
so that the net current exchanged at time $t$ 
between the reservoirs and the site $j$ is given by $I_{j} (t)  =  I_{in,j} ( t ) - I_{out,j} ( t )$.
In addition, we also need to derive the average value of the current flowing between two connected sites of 
the graph, say $j$ and $k$, $I_{j,k}$. This is given by 

\beq
I_{j,k} ( t ) = - i J_{j,k}   {\cal I}_{j,k} ( t )  + {\rm c.c.} 
\:\;\;\; , 
\label{curave.2}
\eneq
\noindent
with ${\cal I}_{j,k} ( t ) = {\rm Tr} [c_j^\dagger c_k  \rho ( t )]$ and  with ${\rm c.c.}$ denoting the complex conjugate.

In principle, solving the full set of Lindblad equations for $n_j ( t )$,  $I_{in,j} (t)$,  $I_{out,j} (t)$, and $I_{j,k} (t)$ would allow us to 
recover the full current pattern over a generic $N$-site graph 
connected to external reservoirs. However, on increasing $N$, solving the Lindblad equations, even numerically, 
becomes soon a pretty formidable task to achieve. Indeed,  we should solve  a hierarchical set
of equations in which the expectation values of any combination of $N$ creation and annihilation operators depends on the 
expectation values of combinations of $N$+2 creation  and annihilation operators. 
In order to exactly describe the system dynamics 
we should in principle compute the evolution of all the matrix elements of the  density matrix, whose dimension is $2^N$. 
Apparently, this becomes soon a hardly accomplishable task, even resorting to a fully numerical 
approach. For this reason, in the following we resort to a MF approximation, by
replacing any occurrence of four-fermion operators with the corresponding approximated expression 
derived by means of a pertinent Hartree-Fock MF decoupling. Eventually, we check the consistency 
of our method with a fully numerical approach \cite{rossini} by comparing the corresponding results in small-size (i.e., $L\leq 16$) 
systems. In particular, we set 
 
\begin{eqnarray}
n_j n_k   & \approx &  n_j ( t ) n_{k}  +  n_k ( t ) n_{j}    \nonumber \\
 & - &  {\cal I}_{j, k} ( t )  c_{k}^{\dagger}c_{j}-  {\cal I}_{k,j} ( t )   c_{j}^{\dagger}c_{k}
 \:\:\:\: . 
 \label{hf.1}
\end{eqnarray}
\noindent
with the first two contributions at the right-hand side of Eq.(\ref{hf.1}) corresponding to the Hartree terms, the second 
two contributions to the Fock ones. An important observation to ground the validity of Eq.(\ref{hf.1}) is that we are
assuming an over-all repulsive interaction between fermions. This rules out  the possibility of p-wave superconducting pairing (anomalous) correlations,
which would otherwise have to be accounted for by adding the corresponding anomalous (pairing) term to the right-hand 
side of Eq.(\ref{hf.1}). 

Going through the MF decoupling, we trade $H$ for the corresponding MF Hamiltonian $\bar{H} ( t ) $, 
given by

\begin{eqnarray}
\bar{H} ( t ) &=& \sum_{ j , k = 1}^N \: c_j^\dagger \: \{ [ {\cal H}_0 ]_{j, k } + [ {\cal H}_U ( t ) ]_{j, k } \} \;  c_k 
\nonumber \\
&\equiv& \sum_{ j , k = 1}^N \: c_j^\dagger \: [ \bar{\cal H} ( t )  ]_{j , k } \: c_k 
\;\;\;\; , 
\label{eq:with_interaction}
\end{eqnarray}
\noindent
with

\begin{eqnarray}
 [ {\cal H}_0 ]_{j, k } &=& - J_{j,k} - \mu_j \delta_{j , k } \nonumber \\
  [ {\cal H}_U ( t ) ]_{j, k } &=&  - U_{j,k} {\cal I}_{j,k} ( t )  + 
  \delta_{j,k}\sum_{i=1}^{N} U_{j,i} n_i ( t ) 
  \:\:\:\: . 
 \label{eq:with_interaction2}
\end{eqnarray}
\noindent
From Eqs.(\ref{eq:with_interaction},\ref{eq:with_interaction2}), we see that, within MF approximation, $H$ is traded for the   bilinear Hamiltonian 
$\bar{H} ( t ) $,  with effective time-dependent parameters  determined by the time evolution
of the system. As a result, the LE becomes nonlinear. In particular,   for our system made of  $N$ interconnected points, it  can   
be written in matrix form as

\beq
\dot{\hat{C}} ( t )  =i [\bar{\mathcal{H}}^{t} ( t )   ,\hat{C} ( t )  ]+\hat{\Gamma}-
\frac{1}{2} \{  (\hat{\Gamma}+\hat{\gamma} ),\hat{C} ( t )  \} 
\;\;\;\; , 
\label{eq:HF-master}
\eneq
with the matrix $\bar{\cal H} ( t ) $ introduced in Eq.(\ref{eq:with_interaction}), the 
 bilinear expectation matrix elements  $  [\hat{C} ( t ) ]_{i,j}= n_i ( t ) \delta_{i,j} + {\cal I}_{i,j} ( t ) 
 [ 1 - \delta_{i,j} ]  $, 
and the system-bath coupling matrix elements $ [\hat{\Gamma} ]_{i,j}=\delta_{i,j}\Gamma_{i}$
and $ [ \hat{\gamma} ]_{i,j}=\delta_{i,j}\gamma_{i}$.

Using Eq.(\ref{eq:HF-master}) we can describe a generic system, noninteracting, as well as interacting (in this latter 
case within MF approximation). Letting the system evolve with $t$, it asymptotically 
flows to a NESS, which we determine from the condition $\dot{\hat{C}} ( t ) = 0$. 
In particular, in the noninteracting case, we can find analytical  solutions for the $\hat{C}$ matrix 
characterizing the NESS,    $\hat{C}_*$, 
by imposing     $\dot{\hat{C}}_* = 0$ in Eq.(\ref{eq:with_interaction}). In order to present the solutions in 
a simple, compact form, we define the column vectors $\hat{C}_{f} $ and $\hat{\Gamma}_{f}$ by ``flattening'' 
the tensors $\hat{C}$ and $\hat{\Gamma}$, that is, by setting 
$\hat{C}_{f} \equiv ( [ \hat[C]_{1,1} , \ldots , [ \hat{C}]_{1,N} , \ldots [\hat{C}]_{2,N} , \ldots )^t$ and 
by analogously defining $\hat{\Gamma}_{f}$. As a result, we find

\beq
[ \hat{C}_* ]_{f}=[ \hat{M}_{1}-\hat{M}_{2} ]^{-1} \hat{\Gamma}_{f}
\;\;\;\; , 
\label{fp.1}
\eneq
\noindent
with  the $N^2 \times N^2$ matrices  $\hat{M}_{1}$ and $\hat{M}_{2}$  defined as  

\begin{eqnarray}
\hat{M}_{1} & = &\left\{  i [ {\cal H}_{0} ]^{t}-\frac{1}{2} [ \hat{\Gamma}+\hat{\gamma} ] \right\} \otimes {\bf I}_{N \times N} \nonumber \\
\hat{M}_{2} & =& {\bf I}_{N \times N}  \otimes \left\{ i [ {\cal H}_{0} ] +\frac{1}{2} [ \hat{\Gamma}+\hat{\gamma} ] \right\} 
\:\:\:\: , 
\label{eq:ana}
\end{eqnarray}
\noindent
with ${\bf I}_{N \times N} $ being the $N \times N$ identity matrix. 

In the interacting case, as we discuss above, resorting to the MF approximation, induces nonlinearities in Eqs.(\ref{eq:HF-master}),
resulting in a much   richer set of   possible  NESSs,  depending on the values of the system parameters.
Apparently, in this case Eq.(\ref{fp.1}) does no longer apply and,   in order to find the corresponding fixed points, we have to resort to a fully numerical approach. 

In the following,  we apply Eq.(\ref{eq:HF-master}) to different systems of physical interest, both in the noninteracting, as well
as in the interacting case.

\section{One-dimensional noninteracting chain}
\label{chain}

As a first application of the LE introduced in the previous Section, 
we now study a single, $L$-site fermionic chain in the noninteracting limit. Following the notation 
introduced in Eq.(\ref{eq:deltastar-H}), we set $J_{j,k} = J$ if $j,k$ label nearest neighboring sites of the chain, 
0 otherwise, $\mu_j = \mu$, that is, constant chemical potential, independent of $j$, and $U_{j,k} = 0$
$\forall j,k$. Accordingly, the chain Hamiltonian $H_c$ is given by

\beq
H_{c}=-J
\sum_{j=1}^{L-1} \{ c_{j}^{\dagger}c_{j+1}+c_{j+1}^{\dagger}c_{j}  \}-\mu\sum_{j=1}^{L}c_{j}^{\dagger}c_{j} 
\:\:\:\: . 
\label{eq:chain_H}
\eneq
\noindent
We assume that the chain  is coupled to two reservoirs  at its endpoints corresponding to the sites  $j=1$ and $j=L$.
Both reservoirs can inject electrons into the chain and absorb electrons from the chain. Therefore, 
the coupling between the chain and the reservoirs is described by a total of four, in principle independent, coupling 
strengths,  $\Gamma_1$, $\gamma_1$, $\Gamma_L$ and $\gamma_L$.
When recovering the above couplings from the microscopic theory, we see that they can be expressed in terms 
of the  Fermi distribution function  at the chemical potential of the reservoir, $f$ and of the reservoir  spectral density
 at the chemical potential of the reservoir, $g$. Specifically, we  obtain \cite{petruccione,zoller}

\begin{eqnarray}
\Gamma_{i} &=&g_{i}f_{i}\nonumber \\
\gamma_{i} &=& g_{i}(1-f_{i})
\;\;\;\; , 
\label{rates.1}
\end{eqnarray}
\noindent
with (labeling each reservoir with the index of the site it is connected to) $i=1,L$. 

As paradigmatic regimes, we consider  the symmetric driving, corresponding to  
$g_1=g_L=g$, $f_1=\frac{1}{2}\left(1+f\right)$ and $f_L=\frac{1}{2}\left(1-f\right)$,  and the large bias regime, 
corresponding to $f_1=1$ and $f_L=0$.  In the symmetric case we parametrize  the reservoirs in terms  of the overall coupling 
$g$ and of the  difference $f=f_1-f_L$ (with, assuming, without loss of generality, $f_1 \geq f_L$, 
$0 \leq f \leq 1$).   $f\approx 0$ corresponds to  the linear response regime.  In the large bias limit, 
the system is driven to the out-of-equilibrium regime, in which the  reservoir coupled to site $1$  acts as an electron ``source'', 
by only injecting electrons in the chain from the reservoir, and the reservoir   coupled to site $L$ 
acts as an electron ``drain'',  by only  absorbing electrons from the system. As a result, 
electrons enter the chain at site 1 and must travel all the way down to site $L$, in order 
to be able to exit the chain. Accordingly, the  boundary dynamics is determined only 
by the coupling strengths $\Gamma_1$ and $\gamma_L$, while the bulk dynamics only depends on the hopping strength $J$.

\begin{figure}
\center
\includegraphics*[width=1.1 \linewidth]{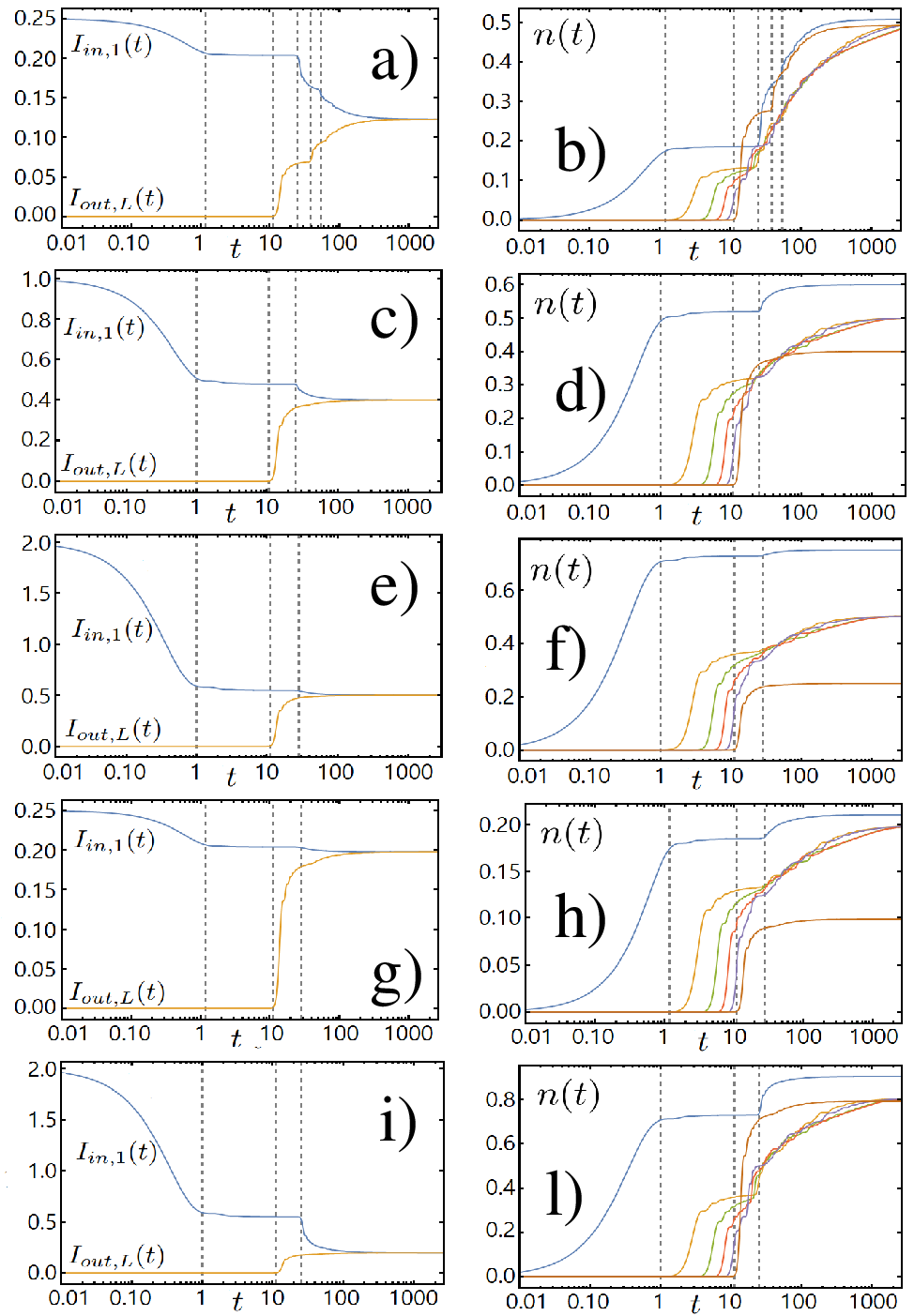}
\caption{ \\ {\bf a)}: $I_{1,in} ( t)$ (blue curve) and $I_{out,L} ( t )$ (yellow curve) currents  as a function of 
time $t$ (on a logarithmic scale) measured in units of $J^{-1}$, in an $L=25$ chain with $J=1$ and $\mu = 0$,  
taken to the large bias  regime for  $ ( \Gamma_1=0.25,\gamma_L=0.25  )$; \\
{\bf b)} $n_1 ( t)$ (blue curve), $n_5 ( t)$ (yellow curve), 
$n_{10} (t)$ (green curve), $n_{15} ( t)$ (red curve), $n_{20} ( t)$ (purple curve), and $n_{25} ( t)$ (orange curve) computed 
in the same chain as we used to draw {\bf a)}.
The vertical dashed lines mark the boundaries of the plateaus corresponding to quasistationary
NESSs (see the main text for the discussion of this point); \\
{\bf c)} Same as in {\bf a)}, but with 
$ ( \Gamma_1=1,\gamma_L=1  )$;\\
{\bf d)} Same as in {\bf b)}, but with 
$ ( \Gamma_1=1,\gamma_L=1  )$;\\
{\bf e)} Same as in {\bf a)}, but with 
$ ( \Gamma_1=2,\gamma_L=2  )$; \\
{\bf f)} Same as in {\bf b)}, but with 
$ ( \Gamma_1=2,\gamma_L=2  )$; \\
{\bf g)}  Same as in {\bf a)}, but with 
$ ( \Gamma_1=0.25,\gamma_L=2  )$; \\
{\bf h)}  Same as in {\bf b)}, but with 
$ ( \Gamma_1=0.25,\gamma_L=2  )$; \\
{\bf i)}  Same as in {\bf a)}, but with 
$ ( \Gamma_1=2,\gamma_L=0.25  )$; \\
{\bf l)}  Same as in {\bf b)}, but with 
$ ( \Gamma_1=2,\gamma_L=0.25  )$. All the plots are drawn by setting  $J=1$ and $\mu=0$.    
}
\label{bounce}
\end{figure}
\noindent
Due to the asymmetric role played by the couplings between the chain and   
the reservoirs, we see that, even in the absence of interaction, our system    can reach a  
 NESS, provided one waits a long enough time (note that, in the case of symmetric 
couplings, in order for the system to reach a NESS in a finite time one has to have a nonzero
interaction \cite{rossini}). To evidence this point, in  Fig.\ref{bounce} we draw 
$I_{in,1}(t)$ and $I_{out,L} (t)$, as well as $n_j(t)$ at selected lattice sites, 
in a noninteracting $L=25$ chain with $J=1$ and $\mu=0$ connected to two reservoirs, with the parameters selected as discussed above. In
particular, in   Fig.\ref{bounce}{\bf a), c), e), g), i)}, we 
draw both $I_{in,1}(t)$ (blue curve) and $I_{out,L} (t)$ (yellow curve) as a function of time (measured in units of 
$J^{-1}$), by initializing our system at $t=0$ and by assuming that  $ n_j ( t = 0) = 0$ $\forall j$. 
Moving from plot to plot, we   change the couplings to the reservoirs, $\Gamma_1 , \gamma_L$, as 
detailed in the figure caption. By synoptically looking at all the plots, we note the important common feature that,
whatever the values of $\Gamma_1$ and of  $\gamma_L$ are, the chain always reaches a NESS at a finite time 
$t_{\rm NESS}$, corresponding at the point where the blue and the yellow curves merge into each other. 
In addition to $t_{\rm NESS}$, we also evidence other (preceding) values of $t$ at which the currents reach 
values that keep stationary  for pretty large intervals of time. 

To physically interpretate the onset of the plateaus in the current, in Fig.\ref{bounce}{\bf b), d), f), h), l)}
we display $n_j ( t )$  at selected sites of the chain as a function of $t$. 
In particular, in each plot we show $n_1 (t)$ (blue curve), $n_5 (t)$ (yellow curve), 
$n_{10} ( t )$ (green curve), $n_{15} (t)$ (red curve), $n_{20} ( t )$ (purple curve), and $n_{25} (t)$ (orange curve). The   values of 
$\Gamma_1$, $\gamma_L$ are the same as for the corresponding plots at the left-hand side (see figure caption for 
details). Over all,  we observe a similar qualitative behavior for all the plots, with $\Gamma_1$ and  $\gamma_L$ only affecting  numerical values
of the various quantities. At the start, we see that electrons enter from site  $j=1$ and start to fill the chain by 
propagating to the right. During this ``pure filling'' phase all the local densities increase in time, in decreasing order,  
from the one corresponding to the leftmost site ($j=1$).  In particular, at   about $t=1$ (marked by the leftmost dashed vertical line in the plots), 
the entering  current and density at the boundary site  $j=1$  reach ``quasistationary'' values that keep constant for a large interval 
of values of $t$ and  only depend  on $\Gamma_1$ (and on $J$, of course). These values correspond to what would be the ``true'' NESS solution 
in  the thermodynamic limit,  $L\rightarrow\infty$. Instead, in our chain the finite size effects determine a breakdown of the NESS above. Indeed,
at $t\sim 10$ (second dashed vertical line from the left), electrons have had enough time to reach the endpoint of the chain 
opposite to the injection point. Accordingly, the density  at $j=L$   starts to grow. At the same time, the outgoing current increases with a slope depending on  $\gamma_L$.
However not all the electrons reaching the endpoint of the chain exit to the right-hand reservoir: a finite  fraction of them is backscattered toward the left-hand endpoint. 
This gives rise to a ``countercurrent'', flowing from the right to the left, and to a corresponding further increase of the local densities, this time 
in reverse order ($j=L$ first). The countercurrent and the further increase of the local densities are exactly the features that take 
the system out of the first putative NESS to a second putative NESS, corresponding to the second shorter plateau in the plots of 
Fig.\ref{bounce}. They are a consequence of having a finite-$L$ chain and, as we argue above, are expected to disappear 
as $L \to \infty$, where the putative NESS becomes the actual NESS of the system. 

Going further ahead in time, at   $t \sim 28$ 
(third dashed vertical line from the left), the countercurrent hits the left-hand endpoint of the chain, 
with the effect of further increasing  $n_1(t)$ and of reducing the incoming current. At this point, the second NESS breaks down, as well. Going ahead in 
time, we see that   electrons   propagate back and forth inside the chain, with a series of consecutive
bounces that manifest themselves in the plots as a series of steps and plateaus,  till the system reaches the ``true'', asymptotic  NESS. The number, the size and the distance of the steps,
as well as the details of the asymptotic NESS,  depend in a non trivial way on  $\Gamma_1$ and  $\gamma_L$.

Looking at the plots in Fig.\ref{bounce} we readily note that the NESS is characterized by 
the convergence (in time) of $I_{in,1} ( t)$ and of $I_{out, L} ( t )$ towards a single, 
time independent, value of the current, $I_{\rm st}$, that is a typical feature of the 
stationary state. Moreover, we also see that $n_j ( t)$ at any site $j \neq 1,L$ flows 
toward a unique value $n_{\rm st}$, thus yielding a profile of the real space 
electron density in the NESS, $n_{{\rm st} , j}$, constant everywhere but at the endpoints of 
the chain. Whether the system is interacting, or not, throughout all the paper 
we characterize the NESS in terms of $I_{\rm st}$ and of $n_{{\rm st} , j}$, in 
particular referring to $n_{\rm st}$ in the flat part of the density profile. 
In the noninteracting case, both $I_{\rm st}$ and $n_{\rm st}$ can be 
analytically determined.   Specifically, 
using   Eq.(\ref{eq:ana}), we obtain

\begin{eqnarray}
I_{\rm st} &=& \lim_{t\rightarrow\infty} \{ \Gamma_{1} (1-n_{1} (t) )-\gamma_{1} n_{1} ( t ) \} \label{asyness.1} \\ &=& 
 \lim_{t\rightarrow\infty} \{ -\Gamma_{L} (1-n_{L} ( t ) )+\gamma_{L}n_{L} ( t ) \}   \nonumber \\ 
&=&\frac{4J^{2} (\Gamma_{1}\gamma_{L}-\gamma_{1}\Gamma_{L} )}{ (\Gamma_{1}+\gamma_{1}+\Gamma_{L}+\gamma_{L} ) [4J^{2}+ (\Gamma_{1}+\gamma_{1} )  (\Gamma_{L}+\gamma_{L} ) ]}
\:\:\:\: . 
\nonumber
\end{eqnarray}
\noindent
From Eq.(\ref{asyness.1}), we see that, as it must be,  $I_{\rm st}$ 
 is  independent of  $L$: the length of the chain only affects the time windows corresponding to each  
 putative NESS.  Similarly, defining $n_{\rm st}$ as 
 
\beq
n_{\rm st}  =\lim_{t\rightarrow\infty} \frac{1}{L-2} \sum_{i=2}^{L-1} \: n_{i} ( t )  
\:\:\:\: , 
\label{asyness.2}
\eneq
\noindent
we obtain 

\begin{eqnarray}
&& n_{\rm st} = \label{asyness.3} \\
&& \frac{4J^{2} (\Gamma_{1}+\Gamma_{L} )+\gamma_{1}^2 \Gamma_{L}+\Gamma_{1} [ (2\gamma_{1}+\Gamma_{1} )
\Gamma_{L}+ (\gamma_{L}+\Gamma_{L} )^2 ]}{ (\Gamma_{1}+\gamma_{1}+\Gamma_{L}+\gamma_{L} )
 [4J^{2}+ (\Gamma_{1}+\gamma_{1} ) (\Gamma_{L}+\gamma_{L} ) ]} \nonumber 
 \:\:\:\: ,  
\end{eqnarray}
\noindent
which, as expected, is independent of $L$, as well. 

From Eq.(\ref{asyness.3}) we see that $n_{\rm st} = 1/2$ for any values of $f$ and $g$ in the symmetric regime and for $\Gamma_{1}=\gamma_{L}$ in the large bias regime.
At variance, at  generic values of  $\Gamma_{1}$ and $\gamma_{L}$ and in the large bias regime, we find 
 
\beq
n_{\rm st} =\frac{\Gamma_{1} (4J^{2}+\gamma_{L}^2  )}{ (\Gamma_{1}+\gamma_{L} ) (4J^{2}+\Gamma_{1}\gamma_{L} )}
\:\:\:\: . 
\label{asyness.4}
\eneq
\noindent
 
\begin{figure}
\center
\includegraphics*[width=1 \linewidth]{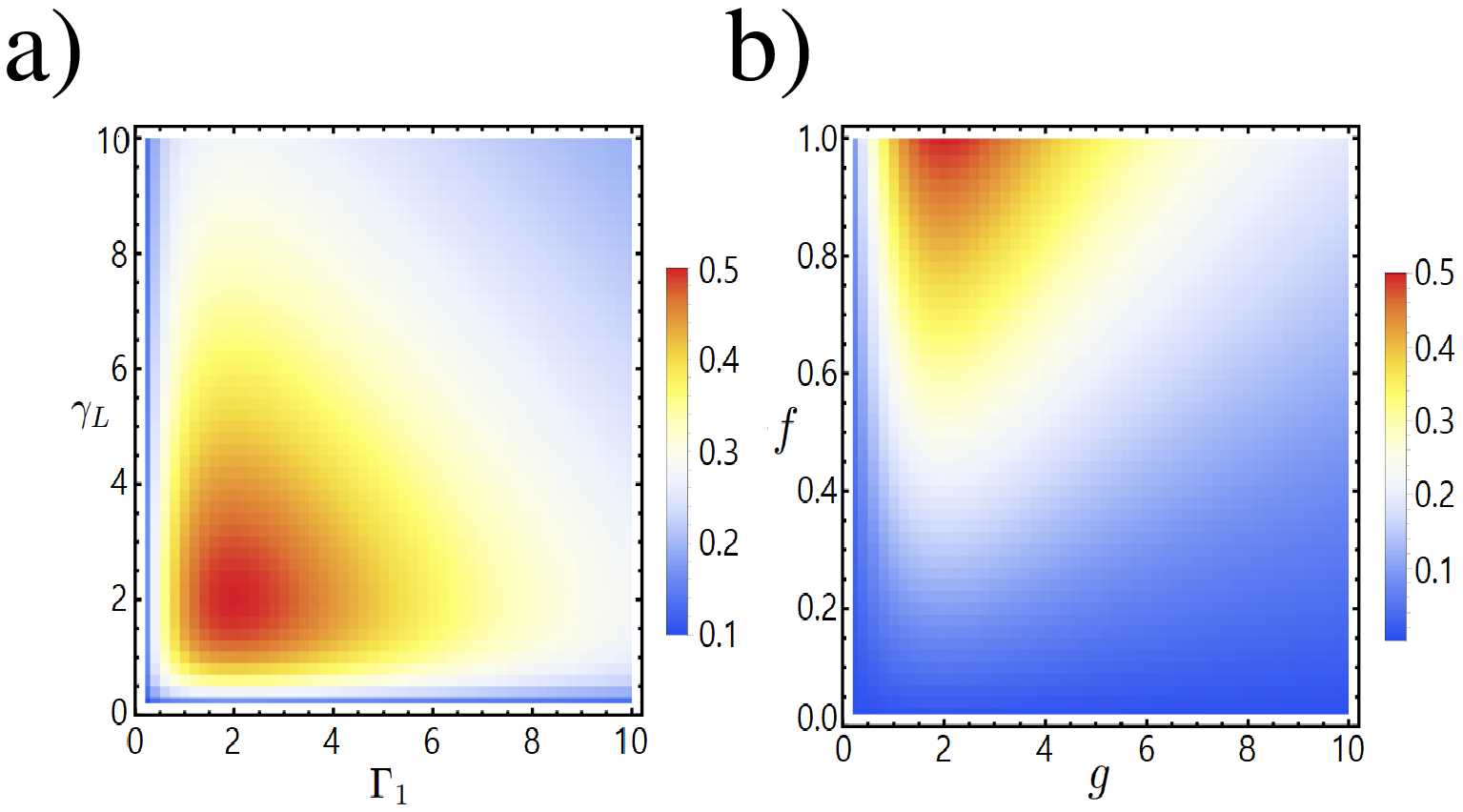}
\caption{\\{\bf a)}: 
$I_{\rm st}$ computed in the $L=25$ chain for $J=1$ and $\mu = 0$  both as a function of $\Gamma_1$ and $\gamma_L$ in the large bias limit; \\
{\bf b)}: $I_{\rm st}$ 
 as a function of $f$ and $g$ in the symmetric regime $\Gamma_1 = \gamma_L $. Red areas correspond to high values of $I_{\rm st}$, blue
areas to low values (see the color code for details).  }
\label{fig:eq_current}
\end{figure}
\noindent

\begin{figure}
\center
\includegraphics*[width=0.6 \linewidth]{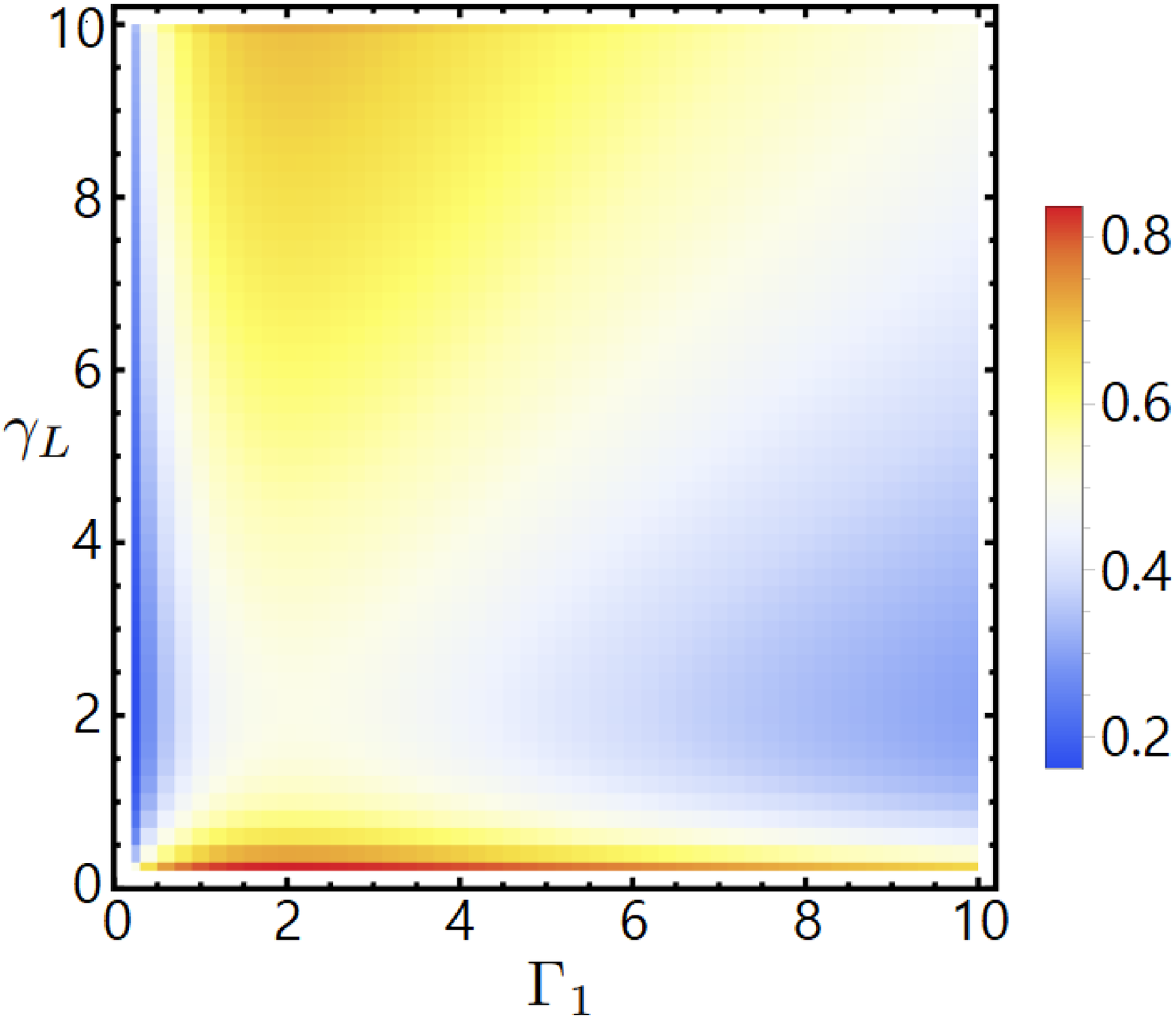}
\caption{$n_{\rm st}$ computed in the $L=25$ chain for $J=1$ and $\mu = 0$   as a function of $\Gamma_1$ and $\gamma_L$ in the large bias limit. 
Red areas correspond to high values of $n_{\rm st}$, blue
areas to low values (see the color code for details).}
\label{fig:density_lb}
\end{figure}
\noindent
To highlight the main features of our chain in the absence of interaction, 
in Fig.\ref{fig:eq_current} we plot $I_{\rm st}$   both as a function of $\Gamma_1$ and $\gamma_L$ in the large bias limit (Fig.\ref{fig:eq_current}{\bf a)}) , 
and as a function of $f$ and $g$ in the symmetric regime $\Gamma_1 = \gamma_L $ (Fig.\ref{fig:eq_current}{\bf b)}) , while in Fig.\ref{fig:density_lb} we plot $n_{\rm st}$
in the large bias regime. Two interesting features emerge. First, we observe the emergence of an OWP in the 
$\Gamma_1 - \gamma_L$ parameter space at which $I_{\rm st}$ is maximum. Specifically, from    Fig.\ref{fig:eq_current}
we see that  the OWP corresponds to  $f=1$ (largest possible bias), and to  symmetric   couplings, 
$\Gamma_{1,max}=\gamma_{L,max}=g_{max}=2$. Second, synoptically considering $I_{\rm st}$  in Fig.\ref{fig:eq_current} and $n_{\rm st}$ 
 in Fig.\ref{fig:density_lb}, we note that the former is, in general, a non-monotonic function of the latter. In particular, we 
 see that  for both high- and low-values of $n_{\rm st}$, $I_{\rm st}$   is lower than its maximum value $1/2$. This is 
 a typical behavior found in   fundamental traffic flow diagrams \cite{trafic_1,trafic_2}, where the free flow phase and the congested phases are separated by an optimal value of 
the density, at which the traffic flow (the ``current'') is maximum.  So, from   Figs.\ref{fig:eq_current} and \ref{fig:density_lb} we see  that 
our system might potentially work as a ``quantum simulator''  of the 
fundamental traffic flow diagram. Aside the fascinating correspondence  with traffic flow phase diagram, 
we definitely evidence how pertinently managing for  the  couplings between the chain and the reservoirs may affect in a nontrivial way
the current propagation into the system.

 In order to compare the LE formalism with alternative approaches to nonequilibrium 
transport problems, such as the non-equilibrium Green function, or the Landauer-B\"uttiker  method,
we refer to the detailed analysis of Ref.[\onlinecite{oliveira}]. In particular, we note that,  
while  we do not expect any substantial qualitative difference between the results obtained within 
those two methods and ours, in order to exactly reproduce the  results using the LE method for $U=0$ 
one should generalize the LE to the case of multisite reservoirs, as extensively discussed, and 
rigorously proven, in Ref.[\onlinecite{oliveira}] , where  it is shown how  the expression for the
nonequilibrium steady-state current of a quantum chain coupled to two multisite reservoirs at both
boundaries, each one consisting of $L_{Res}$ sites,  reduces back to the Landauer-B\"uttiker formula 
for the electronic current  through a tunneling junction, in the limit  $L_{Res} \to \infty$ and for
a small voltage bias between the two reservoirs.

We now generalize our discussion to the case in which a nonzero electron interaction turns on in the chain.

\section{One-dimensional spinless Hubbard chain}
\label{interaction}

Turning on a finite on-site interaction strength $U$ in the chain described by 
$H_c$ in Eq.(\ref{eq:chain_H}), we get the Hamiltonian $H_i$ given by

\beq
H_i = - J \sum_{ j = 1}^{L-1} \{ c_j^\dagger c_{ j + 1 } + c_{ j + 1}^\dagger c_j \}
- \mu \sum_{ j = 1}^L c_j^\dagger c_j + U \sum_{ j = 1}^{L-1} n_j n_{j+1} 
\:\:\:\: .
\label{interac.1}
\eneq
\noindent
$H_i$ in Eq.(\ref{interac.1}) is 1HM Hamiltonian over an $L$-site chain \cite{essler}. To 
recover it from the generic $H$ in   Eq.(\ref{eq:deltastar-H}), we simply set  
$N=L$, $[{\cal H}_0]_{j,k} = - J \{ \delta_{j , k+1} ( 1 - \delta_{k,L} )  +
\delta_{ j ,  k-1} ( 1 - \delta_{ k , 1} ) \} - \mu \delta_{j,k}$, and 
$[{\cal H}_U]_{j,k} = U \{ \delta_{j , k+1} ( 1 - \delta_{k,L} )  +
\delta_{ j ,  k-1} ( 1 - \delta_{ k , 1} ) \}$. 

Beside being a paradigmatic model  for one-dimensional correlated  electronic systems, 
  the 1HM also provides an equivalent description of  an $XXZ$ spin-1/2 quantum spin chain, which is 
mapped onto it via the Jordan-Wigner transformation \cite{jwigner}. Therefore, 
by the same token, when connecting the external reservoirs to the 1HM
described by $H_i$ in Eq.(\ref{interac.1}), we also recover a mean to study the spin
current across a quantum spin chain connected to external reservoirs \cite{rossini_0,rossini,rm_1,ultra}.
Moreover, the 1HM has also been shown to effectively describe a one-dimensional lattice model
of interacting bosons in the limit of strong on-site repulsion between bosons, once the 
chemical potential is tuned so to make the states at each site populated with $n$ and $n+1$ bosons 
to be degenerate with each other \cite{grst}. Therefore, we expect our analysis of the 
interacting fermionic chain to be of relevance for all the different 
physical systems effectively described by the 1HM. 

 $H_i$ in Eq.(\ref{interac.1}) is a simple, prototypical example of 
a strongly correlated fermionic model.  While  the LE  
approach can be implemented in order to study the equilibrium properties of
a quantum system, we do not expect that the mean
field approximation works in the equilibrium case for one-dimensional electronic chains.  
Instead, in this regime different methods, such as  the functional renormalization group 
\cite{karra} or the Thermodynamic Bethe Ansatz \cite{taka}, can be used to study the 
phase diagram of the system. Both methods can be successfully extended to nonequilbrium 
systems: the Thermodynamic Bethe Ansatz can be   implemented successfully to 
analyze nonequilibrium homogeneous quantum chains \cite{doyon},  or quantum 
impurity models \cite{annoneq}; nonequilibrium extensions of the functional
renormalization group approach  have been proposed to study the non equilibrium 
properties of a quantum wire coupled to two reservoirs \cite{addi.1,rg_2}.

In one spatial dimension, it is well known 
that, to analytically describe equilibrium physics of the 1HM, one has to resort to 
sophisticated mathematical techniques, such as the bosonic Luttinger liquid (LL)
approach (see, for instance, Ref.[\onlinecite{giamarchi}] for a comprehensive review
on the subject). As LL is basically a low-energy, long-wavelength effective theory for 
the correlated fermionic system, its applications to strongly out-of-equilibrium states,
such as the ones we discuss here, are not straightforward, not even after resorting to 
clever out-of-equilibrium implementation of the method, such as the one based on 
the nonequilibrium functional renormalization group approach \cite{addi.1}. As 
it is out forward in detail in Refs.[\onlinecite{rossini,noneq}] and as we discuss in 
detail in the following, the NESS that sets in in the nonequilibrium chain 
corresponds to a (combination of) highly-excited states of $H_i$ in Eq.(\ref{interac.1}), 
which are expected to be out-of-reach of the standard LL approach. Moreover, while 
nonequilibrium renormalization group approach might in principle be employed to 
recover the NESS in the nonequilibrium 1HM, the unavoidable technical difficulty of 
extending the approach beyond the regime of weak coupling between the reservoirs and 
the system makes it pretty challenging to recover the density profile and the 
current pattern characterizing the NESS. At variance, as we highlight below, our MF
approximation provide a simple analytical mean to access features of the NESS that
are in a good agreement with results recovered within alternative numerical 
methods \cite{rossini,rossini_0,noneq}. 

In general, $U$ can either be positive, or negative. In the following we 
restrict ourselves to the $U \geq 0$ case only. The negative-$U$ 1HM can nevertheless 
be straightforwardly analyzed by methods analogous to the ones we employ here. 
Also, in all our  calculations, we set  $\mu$   according to 
 the condition that, at equilibrium, the chain is half-filled. This implies 
 choosing $\mu$  so to cancel the chemical potential renormalization due to a finite $U$. 
A simple calculation   
provides the condition $\mu_{\rm eff} = 0$, with $\mu_{\rm eff} = \mu + \frac{U}{2}$. 
In the following, we assume that this condition is already satisfied, unless explicitly 
stated otherwise. To analyze the 1HM connected to the external reservoirs,  we systematically 
implement the MF   decoupling of Eq.(\ref{hf.1}). In fact, at the price of using an 
effective, time-dependent Hamiltonian in the LE, the MF massively eases 
the numerical solution of the LE, compared to   fully numerical approaches \cite{rossini_0,rossini,rm_1,ultra,mun},   
thus allowing for exploring pretty large windows of variations of 
the system parameters. 

Applying Eq.(\ref{hf.1}) to the  interaction $U n_j n_{j+1}$, we obtain 
the corresponding MF decoupling 

\begin{eqnarray}
U n_j n_{j+1}  &\to&  U \{  n_j ( t )   n_{j+1} + n_j n_{j+1} ( t )  \nonumber \\ &-& 
{\cal I}_{j+1 , j} ( t ) c_j^\dagger c_{ j + 1} - {\cal I}_{j , j+1} ( t ) c_{j + 1}^\dagger c_j \} 
\;\;\;\; , 
\label{hf.2}
\end{eqnarray}
\noindent
with   the explicit dependence on $t$ of the average values being a direct consequence of 
the nonequilibrium due to the coupling to the reservoirs. In principle, to determine the time evolution of the 
system   within MF approximation, we have to solve Eq.(\ref{eq:lindbladeq}) for 
$\rho ( t )$ using the Lindblad operators in Eqs.(\ref{eq:deltastar-L}) and the time-dependent 
Hamiltonian $\tilde{H}_i ( t )$, given by 

\begin{eqnarray}
&& \tilde{H}_i ( t )  =  \\ \label{hf.3}
&&-  \sum_{ j = 1}^{L-1} \{ [ J + U {\cal I}_{j+1,j} ( t ) ] c_j^\dagger c_{j+1} 
+ [ J + U {\cal I}_{j,j+1} ( t ) ] c_{j+1}^\dagger c_j \} \nonumber \\
&&-  \sum_{ j = 1}^L [ \mu + U n_{j-1}  ( t ) ( 1 - \delta_{j,1} ) + U n_{j+1} ( t ) ( 1 - \delta_{j , L} ) ] 
n_j
\;\:\:\: . 
\nonumber
\end{eqnarray}
\noindent
Nevertheless, $\tilde{H}_i ( t )$ contains time-dependent averages of operators, which require the 
explicit knowledge of the density matrix at time $t$, in order to be computed. For this reason, 
we numerically solve  the {\it nonlinear} ME equations (Eq.(\ref{eq:observable-master equation})),  directly written for the time-dependent 
averages $n_j ( t )$ and ${\cal I}_{j , j\pm 1}( t ) $. Numerically integrating the nonlinear equations
and taking the large-time limit of the final result, we eventually extend to the interacting case 
the characterization of the NESS in terms of  $I_{\rm st}$ and of $n_{{\rm st},j}$.

\begin{figure}
\center
\includegraphics*[width=1  \linewidth]{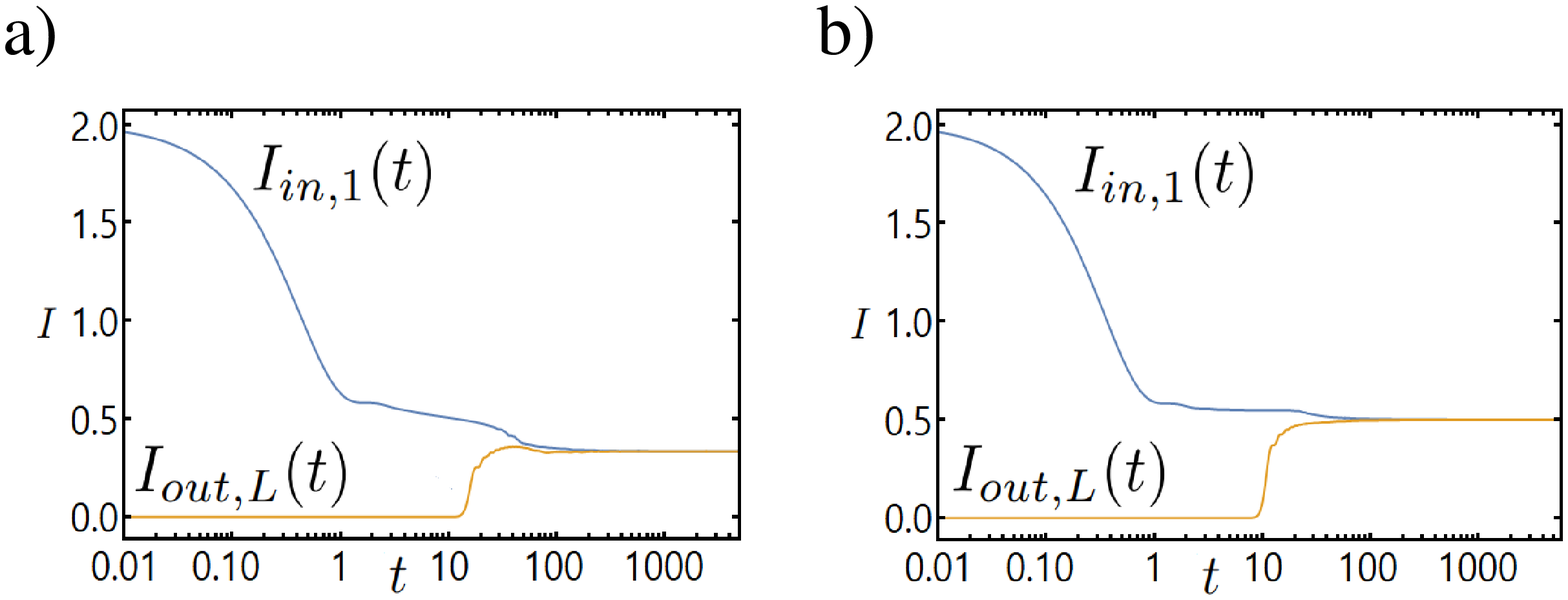}
\caption{  \\ {\bf a)}:   $I_{in,1} (t)$ (blue curve) and  $I_{out,L}(t)$ (yellow curve) currents  as a function of 
time $t$ (on a logarithmic scale) measured in units of $J^{-1}$, in an $L=20$ interacting chain with $U=1$,  
taken to the large bias  regime for   
$   \Gamma_1= \gamma_L=g = 1$,  $J=1$,  and $\mu_{\rm eff} =0$.  \\
{\bf b)}: Same as in {\bf a)}, but with $U=0$.}
\label{curcomp}
\end{figure}

\begin{figure}
\center
\includegraphics*[width=1  \linewidth]{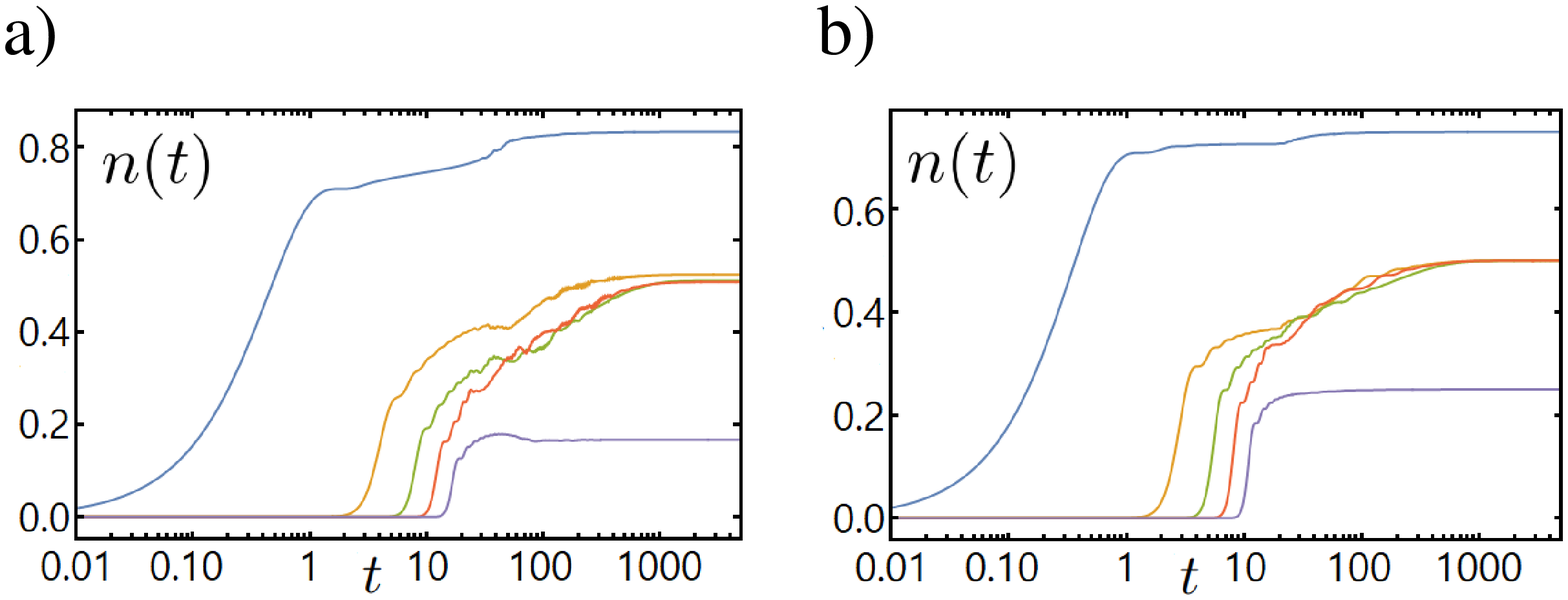}
\caption{\\ {\bf a)}:    $n_1 (t)$ (blue curve), $n_5 (t)$ (yellow curve), $n_{10} (t)$ (brown curve), $n_{15} (t)$ (orange curve),
and $n_{20} (t)$ (purple curve) as a function of 
time $t$ (on a logarithmic scale) measured in units of $J^{-1}$, in an $L=20$ interacting chain with $U=1$,  
taken to the large bias  regime for   
$   \Gamma_1= \gamma_L=g = 1$,   $J=1$,  and $\mu_{\rm eff} =0$; \\  
{\bf b)}: Same as in {\bf a)}, but with $U=0$.}
\label{denscomp}
\end{figure}

In Fig.\ref{curcomp} we plot $I_{in,1} ( t)$ (blue curve) and $I_{out,L} (t)$ (yellow curve)   as a function of 
$t$ (on a logarithmic scale) measured in units of $J^{-1}$, in an $L=20$ chain with $J=1$ and $\mu_{\rm eff} = 0$, in the large-bias regime, 
with $\Gamma_1 = \gamma_L = g = 1$, for $U=1$ (Fig.\ref{curcomp}{\bf a)}), and for $U=0$  (Fig.\ref{curcomp}{\bf b)}). 
In both cases we set the initial state of the chain with $n_j ( t = 0) = 0$ $\forall j$. 
While, from the qualitative point of view, we see no relevant differences between the 
two plots, quantitatively we note a remarkable reduction in 
$I_{\rm st}$ for $U=1$. Such a behavior is known from numerical simulations to emerge in 
the out-of-equilibrium chain, due to the peculiar nonequilibrium  charge density distribution (spin magnetization 
distribution in the corresponing $XXZ$ spin chain) that sets in the system at the NESS \cite{rossini,noneq}. While we
extensively discuss about this point in the following of this Section, we now consider 
  Fig.\ref{denscomp}, where we plot the average particle densities 
at various sites, computed for the same values of the system parameters we used to 
draw Fig.\ref{curcomp}, for  $U=1$ (left-hand plot) and for $U=0$ (right-hand plot).
Specifically, in both plots we draw  
$n_1 (t)$ (blue curve), $n_5 (t)$ (yellow curve), $n_{10} (t)$ (brown curve), $n_{15} (t)$ (orange curve),
and $n_{20} (t)$ (purple curve). 
Apparently, the interacting case is qualitatively similar to the noninteracting one, 
except that it takes a longer time for the sites distant from the current injection point 
to be filled with particles, due to the  repulsive interaction between electrons.

\noindent

\begin{figure}
\center
\includegraphics*[width=1.  \linewidth]{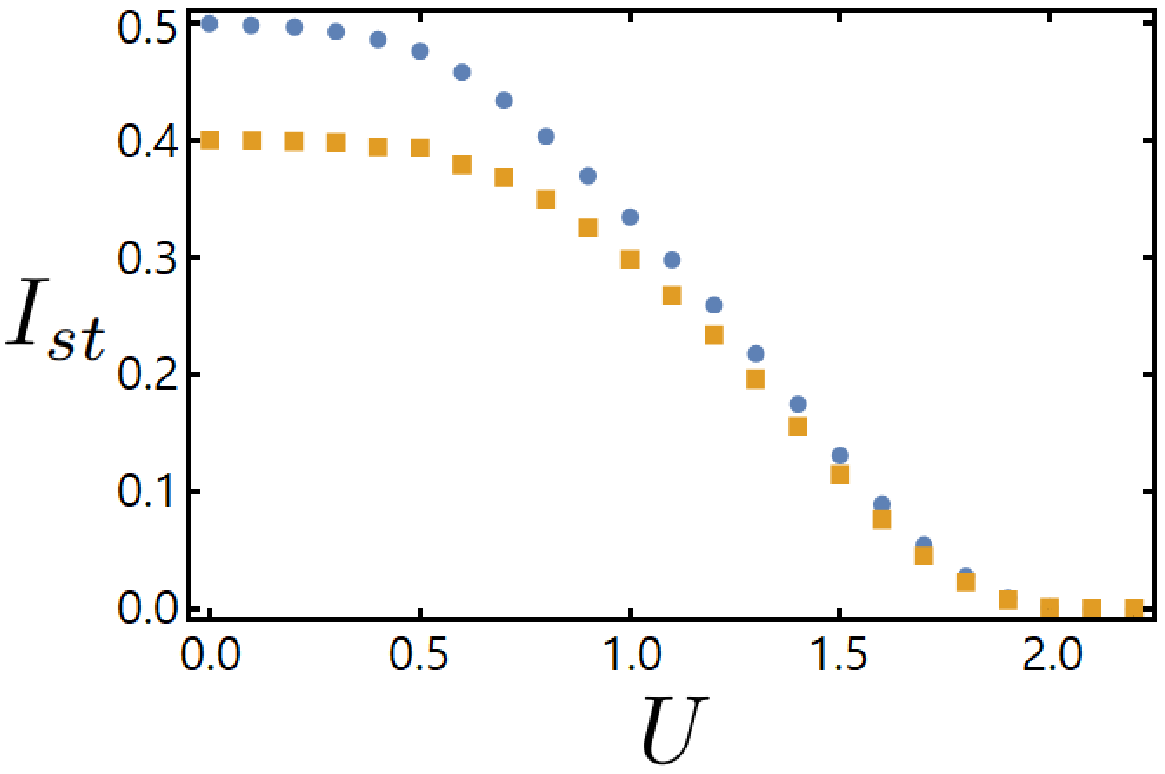}
\caption{$I_{\rm st}$ as a function of $U$ in an $L=20$ chain with $J=1,\mu_{\rm eff} = 0$,
computed in the large bias limit, with $\Gamma_1 = \gamma_L = g = 1$ (orange squares) and with 
 $\Gamma_1 = \gamma_L = g = 2$ (blue circles). There is an apparent conductor to insulator transition 
 at $U = 2$ in both cases.}
\label{iint}
\end{figure}

\noindent
To characterize the NESS, 
we look at the dependence of $I_{\rm st}$ and of $n_{{\rm st} , j}$ on 
the system parameters, starting from the interaction strength $U$.  In Fig.\ref{iint}, we plot 
$I_{\rm st}$  as a function of $U$ for 
specific values of the interaction, ranging from $U=0$ to $U=2$. $I_{\rm st}$  is 
finite, though decreasing with $U$, as long as $\frac{ U }{ J} \leq 2$. At $\frac{U}{J} = 2$, 
$I_{\rm st}$ becomes zero and keeps zero at any $U > 2 J$. Apparently, this is a conductor-to-insulator
transition that, once one goes through the appropriate Jordan-Wigner transformation, is  the analog 
of the behavior of the spin current across an $XXZ$ chain connected to two reservoirs kept 
at large bias, when the Ising anisotropy $\Delta > 1$ (which corresponds to 
$\frac{U}{J} > 2$ in our units) \cite{rossini_0}.  

 About the onset of the insulating phase, 
it is worth pointing out, here, that it is different from the Mott transition toward the insulating 
charge density wave (CDW) phase that takes place at large enough $U$ in the 1HM close to half-filling \cite{giamarchi}. 
Indeed, the CDW sets in as an ordered, staggered pattern in the spatial charge distribution in the equilibrium 
state of the chain. Instead, in the nonequilibrium 1HM we recover a fully different NESS, as we 
discuss in detail next. 

\begin{figure}
\center
\includegraphics*[width=1  \linewidth]{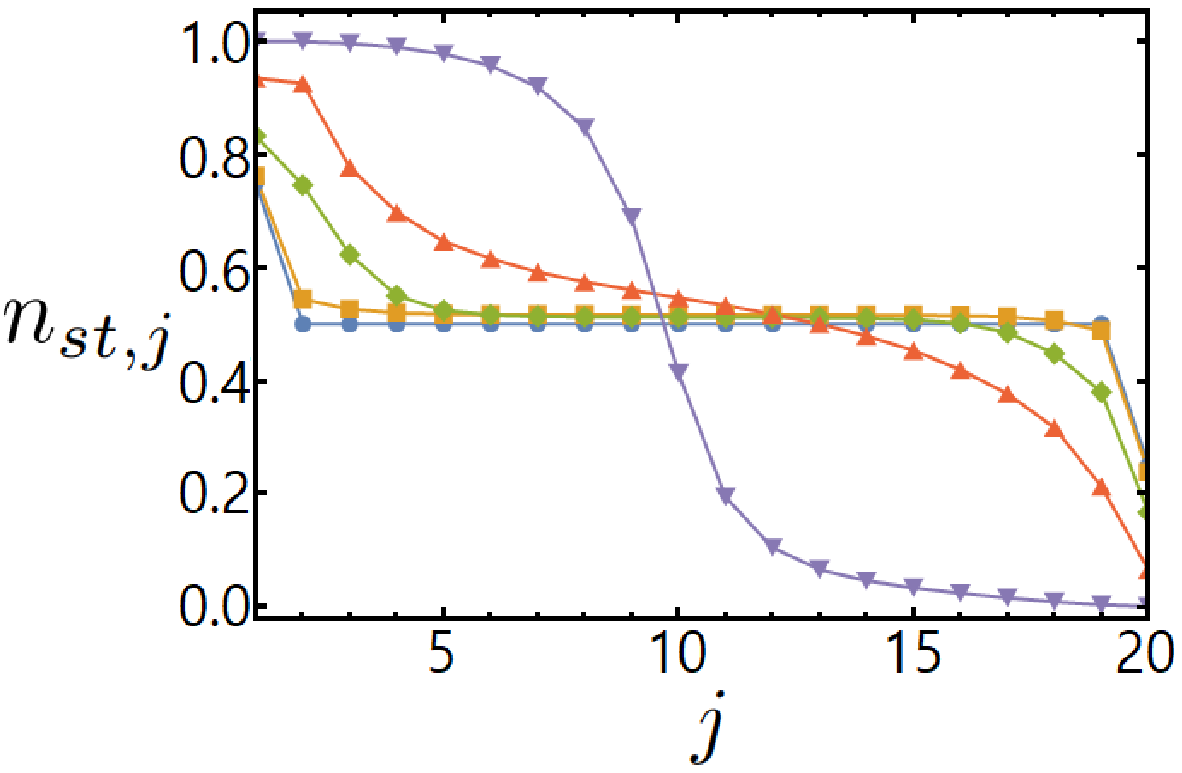}
\caption{ $n_{{\rm st},j}$  at each site $j$ of an $L=20$  chain with $J=1,\mu_{\rm eff} = 0$,
computed in the large bias limit, with $\Gamma_1 = \gamma_L = g = 2$, and with 
$U=0$ (blue circles - blue interpolating curve), $U=0.5$ (orange squares - orange interpolating curve),
$U=1$  (green rhombi - green interpolating curve), $U=1.5$ (red triangles - red interpolating curve), 
and $U=2$ (purple rotated triangles - purple interpolating curve)}
\label{rea_spa_den}
\end{figure}

\noindent
To discuss the nonequilibrium NESS in the interacting model, 
we refer to  Ref.[\onlinecite{rossini_0}], where it is noted how, when connecting a spin-1/2 $XXZ$ spin chain with 
$\Delta > 1$ to two  fully  polarized spin reservoirs with opposite spin polarizations 
(which corresponds to our large bias limit), the reservoirs induce a net magnetization 
within the chain along the same direction of their spin polarization. Being the reservoirs 
oppositely polarized, at strong enough interaction,  a domain wall arises at the center of the chain, where smoothly, 
though rapidly (in real space) the magnetization profile matches the opposite, ``asymptotic''
values (see Fig.3 of  Ref.[\onlinecite{rossini_0}] for details). The formation of the domain 
wall strongly suppresses the spin current across the chain, thus effectively inducing 
a transition between a ``spin conducting'' and a ``spin insulating'' phase. By analogy, 
in our case we expect a charge domain wall to emerge in the real space 
profile of $n_{{\rm st},j}$ at $\frac{U}{J}=2$. To check this point, in Fig.\ref{rea_spa_den} we plot   
$n_{{\rm st},j}$, at each site of an $L=20$  chain with $J=1,\mu_{\rm eff} = 0$,
taken to  the large bias limit, with $\Gamma_1 = \gamma_L = g = 2$, and with 
$U=0$ (blue circles - blue interpolating curve), $U=0.5$ (orange squares - orange interpolating curve),
$U=1$  (green rhombi - green interpolating curve), $U=1.5$ (red triangles - red interpolating curve), 
and $U=2$ (purple rotated triangles - purple interpolating curve).  While the plot at $U=2$ apparently matches the corresponding one, drawn at 
$\Delta = 1$, in Fig.3 of Ref.[\onlinecite{rossini_0}], for $\frac{U}{J} < 2$, we see extended flat 
regions in the profile of  $n_{{\rm st},j}$, eventually bending upward or downward close to the 
endpoints of the chain.

\noindent

In Appendix \ref{varia}, we provide a physical interpretation of the behavior of 
$I_{\rm st}$ and of  $n_{{\rm st},j}$. In particular, resorting to a simple, though qualitatively effective, 
variational method combined a pertinent MF approach to the 1HM   
\cite{saleur}, we argue how, in order for our system  to support a finite $I_{\rm st}$ 
for $\frac{U}{J} \leq 2$,  there has to be a flat region of values in $n_{{\rm st},j}$  throughout the middle part of the chain, with a respectively upward and a downward turn
close to the endpoints of the chain, that are required to match $n_{{\rm st} , 1}$ and $n_{{\rm st} , L}$ as determined by the constancy 
of $I_{\rm st}$. On increasing $U$, the  flat region shortens, till it shrinks at $U = 2$, by   taking  a ``kink-like'' profile, 
with a corresponding blocking of the current transport ($I_{\rm st} = 0$), for 
$U \geq 2 $ \cite{rossini,rossini_0}. Therefore, when characterizing the NESS by looking at $I_{\rm st}$ and of  $n_{{\rm st},j}$, 
we are apparently led to associate the emergence 
of a flat density region in the middle of the chain with a finite value of the stationary state current and, at variance, 
a kink-like profile in the density plot with a blocking of the charge flow, that is, with   $I_{\rm st} = 0$. 
The bending of the flat density profile as $\frac{U}{J}$ increases corresponds to the reduction of 
$I_{\rm st}$ at increasing $U$, which we display  in Fig.\ref{curcomp}.

\begin{figure}
\center
\includegraphics*[width=1. \linewidth]{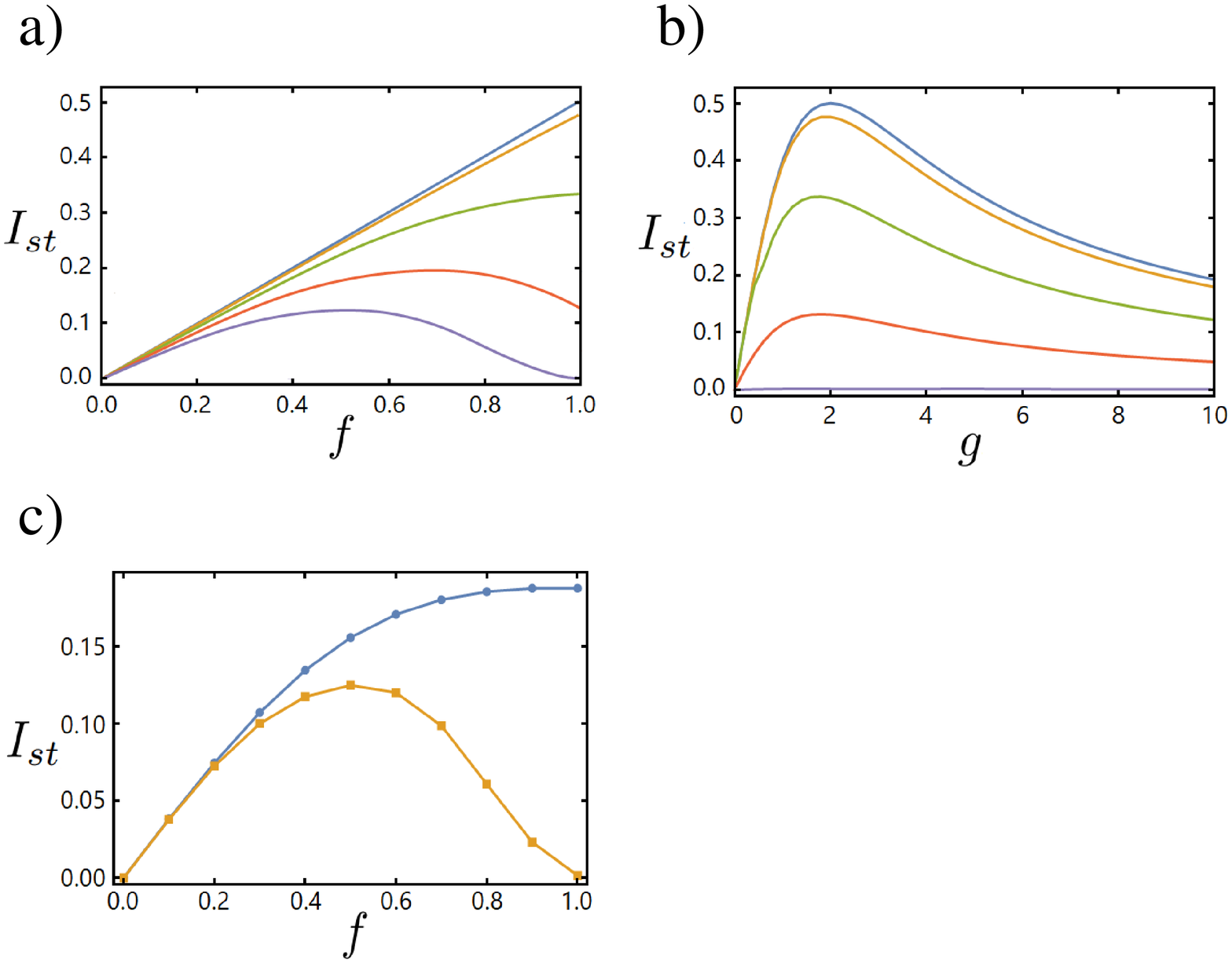}
\caption{\\ {\bf a)}:  $I_{\rm st}$   as a function of $f$, with  $g$ set so that 
$I_{\rm st}$ is maximum, computed in an $L=20$ chain with $J=1$, $\mu_{\rm eff} =0$,
and $U=0$ (blue curve), $U=0.5$ (yellow curve), $U=1$ (green curve), $U=1.5$ (red
curve) and $U=2$ (purple curve); \\
{\bf b)}: Maximum value of $I_{\rm st}$ 
as a function of $g$ in the large bias regime for $g=\Gamma_1=\gamma_L$. The system 
parameters and the color code for the curves drawn at different values of $U$ are
the same as in the left-hand plot; \\
{\bf c)}:  $I_{\rm st}$ as a function of $f$, with  $g$ set so that 
$I_{\rm st}$ is maximum, computed in a chain with $J=1$, $\mu_{\rm eff} =0$, $U=2$, 
with $L=4$(blue dots, blue interpolating curve) and with $L=16$ (yellow squares, yellow interpolating curve).}
\label{fig:interaction_plot}
\end{figure}

To discuss the emergence of the OWP in the extended parameter space including   $U$ and  the coupling strengths, 
  we first, following Ref.[\onlinecite{rossini}], look at the maximum value of 
$I_{\rm st}$ as a function of  $f$ in an $L=20$ chain with $J=1,\mu_{\rm eff} =0$, 
in the symmetric case and for various values of $U$ (Fig.\ref{fig:interaction_plot}{\bf a)}).
Then, we extend the parameter space by considering $I_{\rm st}$ 
 as a function of $g$ in the same chain, in the  large-bias limit, for $\Gamma_1=\gamma_L=g$ (Fig.\ref{fig:interaction_plot}{\bf b)}),
at increasing values of $U$.   
As in Ref.[\onlinecite{rossini}], we find that, at finite $U>0$, an  OWP  
emerges, at a value of $f_{\rm OWP}$ at which $I_{\rm st}$ reaches its maximum value. 
Moreover, the larger is $U$, the more $f_{\rm OWP}$ is pushed towards lower values of $f$.  
In addition, from   Fig.\ref{fig:interaction_plot}{\bf b)}, we also 
recover one the most important original results of our work, that is, that turning 
on $U$ {\it is not} a necessary condition to get the OWP  (see also Appendix 
\ref{varia} for a separate discussion of this point). Indeed, we 
see a maximum in the plots of $I_{\rm st}$ as a function of $g$ even when $U = 0$, 
{\it provided we tune the system at the large-bias limit}. So, we directly prove 
that, increasing the number of tuning parameters, we may recover the OWP even in 
regions in parameter space where it does not emerge if the system is close to the 
equilibrium. More specifically, to make a quantitative comparison with 
the results of  Refs.[\onlinecite{rossini,rossini_0}], we focus onto the purple curve of Fig.\ref{fig:interaction_plot}{\bf a)}. 
Apparently, this exhibits a  reasonable qualitative agreement with the purple (bottom) curve of 
Fig.11 of Ref.[\onlinecite{rossini}], though with  a stronger bending toward the zero-current axis as $L \to 20$, 
which is motivated by the slightly larger  total number of sites (20 rather than 16) and 
by the observation that the system should be insulating in the thermodynamic limit at $\frac{U}{J} = 2$. 
More generally,  the MF approach is expected to underestimate fluctuations in short chains 
and, therefore, to work fine for long enough chains. To check this point, in Fig.\ref{fig:interaction_plot}{\bf c)}
we plot $I_{\rm st}$ as a function of $f$ with the same system parameters used to draw Fig.\ref{fig:interaction_plot}{\bf a)}, 
for $L=4$ (blue dots, blue interpolating curve) and for $L=16$ (yellow squares, yellow 
interpolating curve). We note a better agreement between the corresponding plots of 
Fig.11 of Ref.[\onlinecite{rossini}] relative to the longer chain ($L=16$), rather than the shorter one ($L=4$).

\begin{figure}
\center
\includegraphics*[width=1. \linewidth]{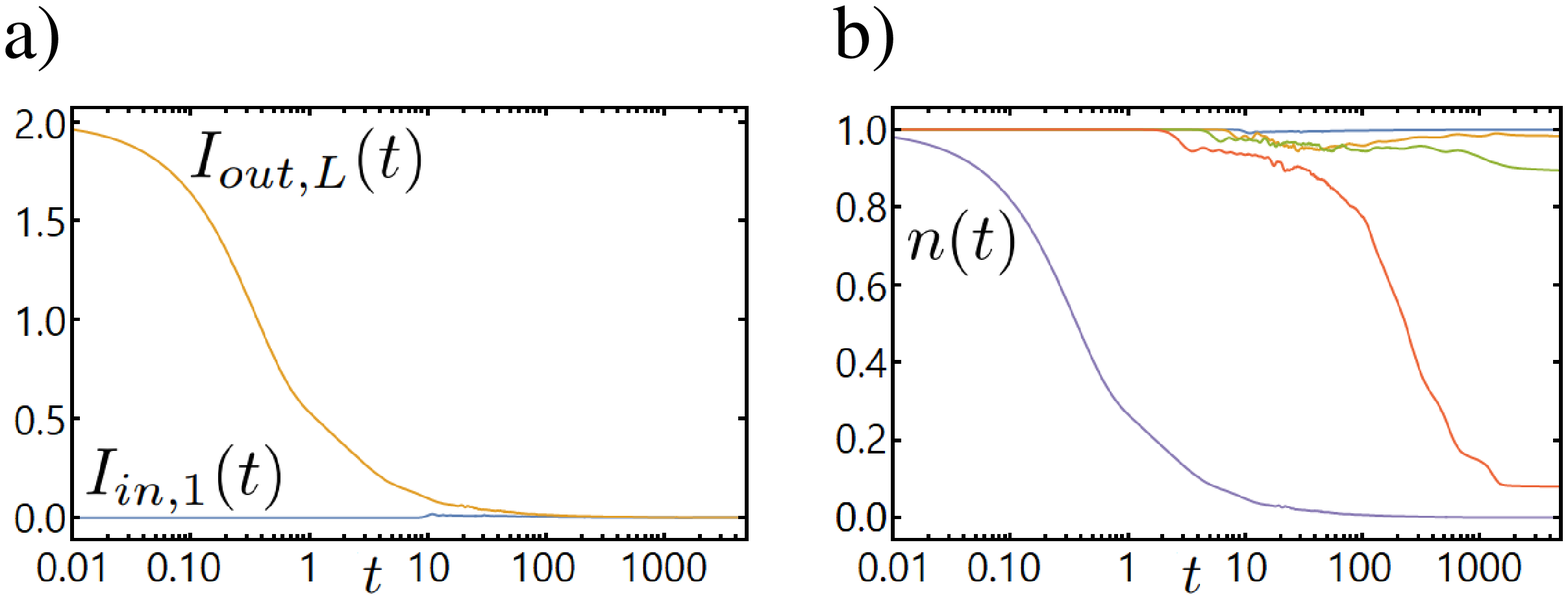}
\caption{\\ {\bf a)}:  $I_{in ,1}( t)$ (blue curve) and $I_{out,L}(t)$ (yellow curve) currents  as a function of 
time $t$ (on a logarithmic scale) measured in units of $J^{-1}$, in an $L=20$ interacting chain with $J=1$, 
$\mu_{\rm eff} = 0$, and $U=2$,  
taken to the large bias  regime for   
$  \Gamma_1=1,\gamma_L= g = 1$,  $J=1$,   with the system prepared, at $t=0$ in the state 
with  $n_j ( t = 0 ) = 1$ $\forall j$; \\
{\bf b)}: $n_1 (t)$ (blue curve), $n_5 (t)$ (yellow curve), $n_{10} (t)$ (green  curve), $n_{15} (t)$ (orange curve),
and $n_{20} (t)$ (purple curve)   as a function of 
time $t$ (on a logarithmic scale) measured in units of $J^{-1}$, in an $L=20$ interacting chain with $J=1$, 
$\mu_{\rm eff} = 0$, and $U=2$,  
taken to the large bias  regime for   
$  \Gamma_1=1,\gamma_L= g = 1$,  $J=1$,   with the system prepared, at $t=0$ in the state 
with  $n_j ( t = 0 ) = 1$ $\forall j$. }
\label{comp_full}
\end{figure}

\noindent
To conclude this Section, we briefly discuss the dependence of the NESS on 
the initial state of the system. This is an important point to verify, 
so to make sure that the NESS is unique and there are no  ``bifurcations'' 
in the time evolution described by Lindblad equations, which 
in some cases may affect the time evolution of the system toward the NESS \cite{lind_5}.
In particular, in  Fig.\ref{comp_full}{\bf a)}
we show $I_{in ,1} (t)$ (blue curve) and $I_{out ,L} (t)$ (yellow curve)    as a function of 
time $t$ in the same system as the ones we have used to derive Figs.\ref{curcomp},\ref{denscomp},
but  with the initial state characterized by $n_j ( t = 0 ) = 1$ $\forall j$. 
 While the evolution in time of the two currents is completely
 different from the one that we report in   Fig.\ref{curcomp}{\bf a)}, we see 
that, as $t \to \infty$, they converge to the same $I_{\rm st}$ as in the case 
in which one has  $n_j ( t = 0 ) = 0$ $\forall j$. For comparison, in  Fig.\ref{comp_full}{\bf b)}, we draw  
$n_1 (t)$ (blue curve), $n_5 (t)$ (yellow curve), $n_{10} (t)$ (green  curve), $n_{15} (t)$ (orange curve),
and $n_{20} (t)$ (purple curve)  as a function of  time $t$. Except for the green curve, 
we find an acceptable consistency with the   values of $n_{{\rm st} , j}$
extrapolated from Fig.\ref{denscomp}{\bf a)}. Being the density a local 
operator, it is strongly affected by the finite size of the chain, by 
the distance from the reservoirs, {\it et cetera}. Likely, working with longer chains 
and extrapolating the results over longer time would wash out this discrepancy, as well. 
So, we may readily infer that, though the initial states in the two cases are completely 
different, 
the NESS in the two systems is the same,  independently of the initial state.

\noindent
We now discuss the effects of nonzero  disorder on the time evolution of our system and on 
the formation of the corresponding NESS.  To do so, we first discuss the case of a single, isolated 
defect in the chain (an ``impurity''). Therefore, we consider a finite density of random impurities in 
the chain (``quenched disorder'').

\section{A single impurity in the chain at large bias}
\label{singleimp}

Impurities in a quantum chain can be either realized by 
tuning  $\mu_{\rm eff}$ at a site $\bar{j}$ to a value $\mu_d$ different from all the other sites (``site impurity''),  or 
  by changing the electronic hopping strength of a single bond of the chain (``bond impurity'').  While 
the equilibrium physics of   impurities in the 1HM  (or in the $XXZ$ spin-1/2 quantum spin chain) can 
be analytically addressed within a number of effective methods, such as LL approach, 
\cite{affleck_1,affleck_2,kanefi_1,kanefi_2,giuso_1,gsta,grt}, it is definitely challenging to 
analytically deal with transport across impurities in the 1HM connected to reservoirs in the large bias 
limit, even after resorting to powerful analytical methods, such as the functional renormalization group
approach developed in Ref.[\onlinecite{addi.1}]. Nevertheless, as we discuss in this Section, despite its simplicity, our MF approach is
able to catch the relevant physics of the NESS state in the out-of-equilibrium 1HM.
 
First of all, we recall that both the   1HM and  $XXZ$ spin chain  are  integrable models. 
In general, the conservation laws associated to integrability 
are known to prevent the system from thermalizing toward a state characterized by a macroscopic hydrodynamical 
behavior in its transport properties \cite{prev_1,prev_2}. On this respect, a crucial problem is analyzing how 
a perturbation breaking the integrability of the system (even locally) affects the evolution toward the NESS in 
the large bias limit \cite{rigol}. In this direction, 
the simplest possible way of breaking integrability is by just making the system inhomogeneous by adding a local 
impurity term to the otherwise integrable Hamiltonian. 

Following the approach of Ref.[\onlinecite{noneq}], in this Section we analyze how adding an impurity 
term to the Hamiltonian in Eq.(\ref{interac.1}) affects the evolution of the system toward the NESS, and 
the structure of the NESS itself, once the chain is connected to the reservoirs out of equilibrium.  For the sake of 
simplicity in both cases, with no loss of generality, we symmetrically realize the impurity at the 
center of the chain, which requires $L$ to be odd for the site impurity and even for the bond impurity.

\begin{figure}
\center
\includegraphics*[width=1. \linewidth]{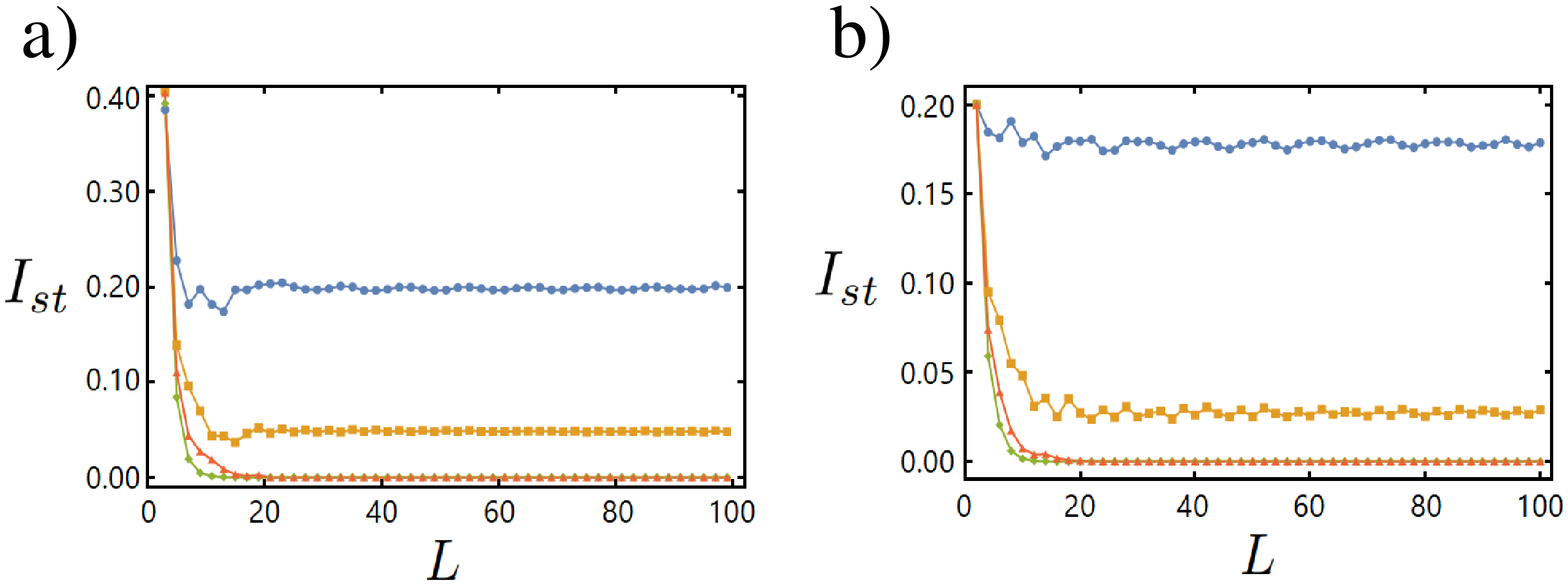}
\caption{\\{\bf a)}: 
$I_{\rm st}$ as a function of $L$ computed in the 1HM with a site impurity of strength  $\mu_d=1.5$, with $J=1$ and 
$U=1.0$ (blue circles, blue interpolating  curve), $U=1.7$ (yellow squares, yellow interpolating curve),
$U=2.0$ (orange triangles, orange interpolating curve), $U=3.0$ (green rhombi, green interpolating curve);  \\
{\bf b)}: $I_{\rm st}$ as a function of $L$ computed in the 1HM with a bond  impurity of strength $J_d=0.5$, with $J=1$ and   
$U=1.0$ (blue circles, blue interpolating  curve), $U=1.7$ (yellow squares, yellow interpolating curve),
$U=2.0$ (orange triangles, orange interpolating curve), $U=3.0$ (green rhombi, green interpolating curve)  }
\label{cur.1}
\end{figure}
\noindent

\begin{figure}
\center
\includegraphics*[width=1. \linewidth]{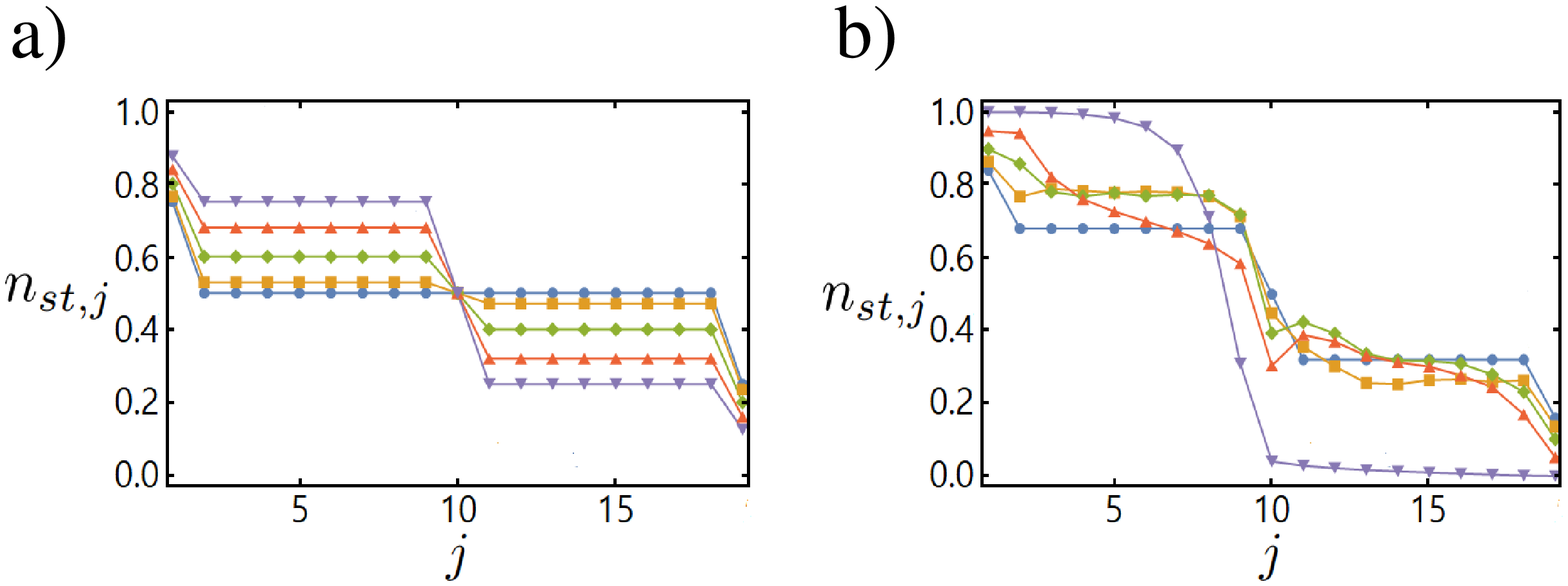}
\caption{\\ {\bf a)}:   $n_{{\rm st},j}$, computed in an HM with $L=19$, $J=1$,
$\mu_{\rm eff} = 0$, and $U=0$,  in the large bias limit with $\Gamma_1 = \gamma_L = g = 2$, with a site corresponding to the Hamiltonian in Eq.(\ref{imp.1}) with 
$\mu_d = 0.0$ (blue dots, blue interpolating curve), $\mu_d = 0.5$ (orange squares, orange interpolating curve), 
$\mu_d = 1.0$ (green rhombi, green interpolating curve), $\mu_d = 1.5$ (red upward pointing triangles, red interpolating 
curve), and $\mu_d = 2.0$ (purple downward pointing triangles, purple interpolating curve); \\ 
{\bf b)}:  $n_{{\rm st},j}$, computed in an HM with $L=19$, $J=1$,
$\mu_{\rm eff} = 0$,  in the large bias limit with $\Gamma_1 = \gamma_L = g = 2$,  with a site corresponding to the Hamiltonian in Eq.(\ref{imp.1}) with 
$\mu_d = 1.5$, computed for $U=0.0$ (blue dots, blue interpolating curve), $U = 0.5$ (orange squares, orange interpolating curve), 
$U = 1.0$ (green rhombi, green interpolating curve), $U = 1.5$ (red upward pointing triangles, red interpolating 
curve), and $U = 2.0$ (purple downward pointing triangles, purple interpolating curve). }
\label{imps_1}
\end{figure}

To check the reliability of our method, in the case 
of the site impurity we compare our results with the analogous ones of Ref.[\onlinecite{noneq}]
obtained within $t$-DMRG approach.   Taking $L$ odd, we realize the site impurity   by adding to $H_i$ in 
Eq.(\ref{interac.1}) the impurity Hamiltonian $H_{\rm site}$ given by 

\beq
H_{\rm site} = - \mu_d n_{\frac{L+1}{2} }  = - \mu_d  c^\dagger_{\frac{L+1}{2} }  c_{\frac{L+1}{2} } 
\:\:\:\: . 
\label{imp.1}
\eneq
\noindent
As we have done in the homogeneous case, we characterize the NESS  by looking at  $I_{\rm st}$ and at $n_{{\rm st},j}$. To highlight the effects of increasing 
the impurity interaction strength, in Fig.\ref{imps_1}{\bf a)}  we show $n_{{\rm st},j}$, computed in an 1HM with $L=19$, $J=1$,  
$\mu_{\rm eff} = 0$, and $U=0$, in the large bias limit with $\Gamma_1 = \gamma_L = g = 2$, with the impurity symmetrically located at site $\bar{j} = 10$, 
at various values of the on-site chemical potential $\mu_d$ (see figure caption for details).
At variance, to evidence the effects of the bulk interaction, in    Fig.\ref{imps_1}{\bf b)}
we again show   $n_{{\rm st},j}$, computed for the same parameters as 
in the left-hand plot, except that now we fix $\mu_d = 1.5$ and vary $U$ from plot to plot 
 (see figure caption for details).  Finally, to make sure that there are no finite-size 
 effects spoiling our results, we compute $I_{\rm st}$ as a function of $L$ for both the site impurity and 
 the bond impurity for selected values of $U$ up to $L \sim 100$, where, as we show 
in  Fig.\ref{cur.1}{\bf a)} for the site impurity and in  Fig.\ref{cur.1}{\bf b)} for the bond 
impurity, apparently $I_{\rm st}$ has reached its asymptotic value in the thermodynamic limit.

\begin{figure}
\center
\includegraphics*[width=1. \linewidth]{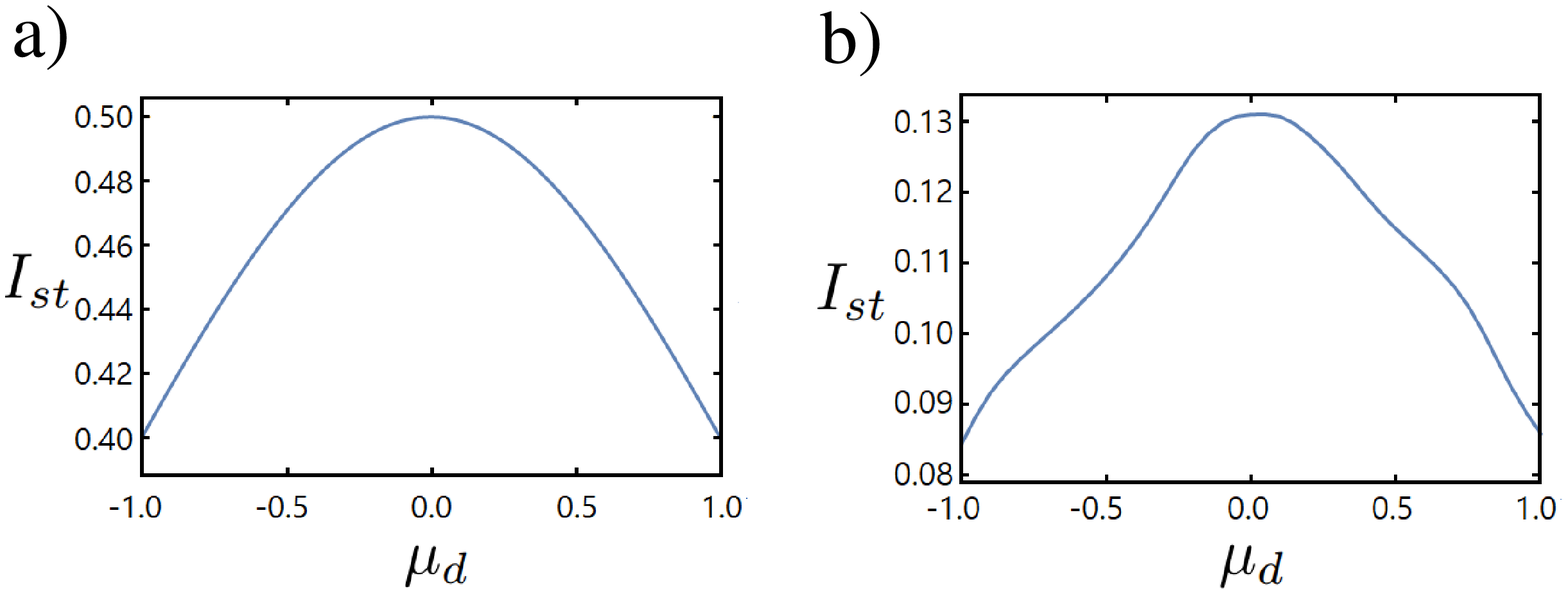}
\caption{\\ {\bf a}: $I_{\rm st}$ as a function of $\mu_d$ computed in an HM with $L=19$, $J=1$,
$\mu_{\rm eff} = 0$, and $U=0$, in the large bias limit with $\Gamma_1 = \gamma_L = g = 2$, 
with a site corresponding to the Hamiltonian in Eq.(\ref{imp.1}); \\ 
{\bf b}: The same as in {\bf a)}, but with $U=1.5$. }
\label{cmps_1}
\end{figure}

\noindent
By synoptically looking at the plots in Fig.\ref{imps_1}, we readily note that a nonzero $\mu_d$ triggers the 
emergence of a kink in the middle of the chain, even for $\frac{U}{J} \leq 2$. This feature appears similar to the 
formation of the kink in the middle of the chain that marks the transition from the conducting to the insulating phase 
of the nonequilibrium 1HM (see Ref.[\onlinecite{rossini_0}] as well as our Appendix \ref{varia} for a detailed 
discussion of this point). However, the persistence of finite regions in real space where the density 
keeps flat, a feature that is associated to the conducting phase of the 1HM \cite{noneq}, is already a 
clue that the NESS in the presence of a site impurity is only quantitatively, not qualitatively, different
from the NESS that sets in the homogenous chain for $\frac{U}{J} \leq 2$. To double check this conclusion, in Fig.\ref{cmps_1} 
we plot $I_{\rm st}$ as a function of $\mu_d$ computed in the same $L=19$ chain we used to derive the 
plots of Fig.\ref{imps_1}, with $J=1$, $\mu_{\rm eff} = 0$, and $U=0$ (Fig.\ref{imps_1}{\bf a)}), and $U=1.5$  (Fig.\ref{imps_1}{\bf b)}),
in the large bias limit with $\Gamma_1 = \gamma_L = g = 2$.
Whether $U = 0$ or $U$ takes a finite value, we see that while,  turning on $\mu_d$   slightly reduces 
$I_{\rm st}$, at the same time the current keeps finite within the NESS. This enforces the conclusion that 
turning on a site impurity in the chain does not qualitatively affect the NESS. Therefore, we conclude that 
both in the case of a homogeneous 1HM, as well as in the presence of a site impurity, our system 
flows toward a conducting NESS, with an extended, flat regions in the profile of $n_{{\rm st},j}$
in the middle of the chain, and with a finite value of  $I_{\rm st}$.  In fact, as discussed in detail in 
Ref.[\onlinecite{noneq}], for this range of values of $\frac{U}{J}$ $I_{\rm st}$ is expected to 
scale with $L$ as $L^{- \nu}$, with $\nu = 0$, that is, what is expected for a ballistic conducting 
channel. Indeed, as we show in  Fig.\ref{cur.1}{\bf a)}, that is what we find within our MF approach, 
with also a reduction in the (uniform) value of $I_{\rm st}$ as $\frac{U}{J}$ increases, that is 
again consistent with Ref.[\onlinecite{noneq}]. For $\frac{U}{J} > 2$, the chain turns into a diffusive 
transport regime, characterized by an exponent $\nu > 0$, corresponding to a suppression of the 
current in the thermodynamic limit. Again, this is consistent with  the result we display in 
 Fig.\ref{cur.1}{\bf a)}. Yet, it is worth stressing that the flow of $I_{\rm st}$ towards its thermodynamic limit, at increasing 
values of $L$, is characterized by short-distance features, such as subleading power-law decays and/or 
small oscillations in the current (in the bond impurity case). We believe that it would be extremely interesting 
to recover analytical formulas for those features, particularly concerning their relation to 
the value of $\frac{U}{J}$ and/or to the impurity strength, an issue that goes beyond that scope 
of this work and which might possibly require a pertinent implementation of sophisticated 
analytical methods, such as the ones discussed in Ref.[\onlinecite{addi.1}].

Concerning the role of integrability and of integrability breaking, we note  how, connecting the homogeneous 1HM to the reservoirs already 
breaks the integrability of the model, thus triggering a flow in real time toward a uniquely defined 
NESS. Indeed, the very fact that adding an additional term breaking the integrability ($H_{\rm site}$) gives rise 
to a feature (the ``central kink'') analogous to the ones arising at the endpoints of the chain 
when it is connected to the reservoirs evidences how in both cases we break integrability, which is consistent 
with our results of Sections \ref{chain},\ref{interaction}.

\begin{figure}
\center
\includegraphics*[width=1. \linewidth]{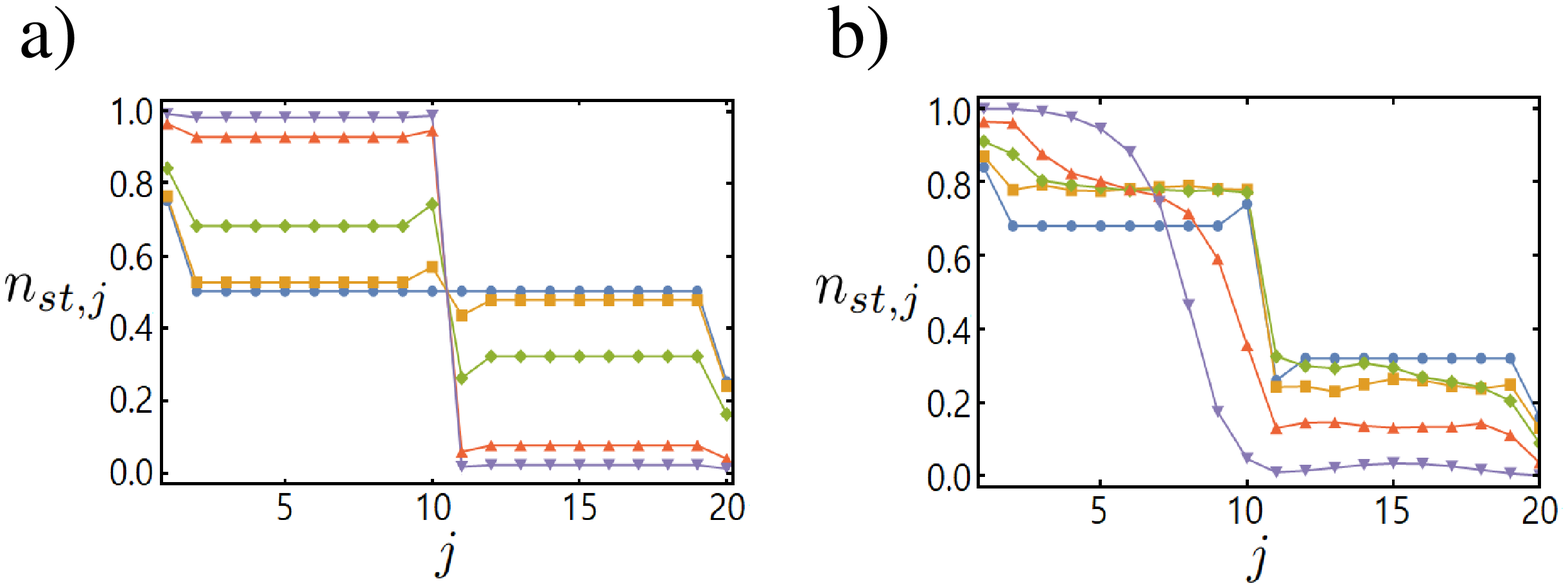}
\caption{\\ {\bf a}:  $n_{{\rm st},j}$ computed in an 1HM with $L=20$, $J=1$,
$\mu_{\rm eff} = 0$, and $U=0$,  in the large bias limit with $\Gamma_1 = \gamma_L = g = 2$, with the bond impurity symmetrically located between sites $\bar{j} = 10$ and 
$\bar{j}+1=11$,  at  $J_d = 1.0$ (blue dots, blue interpolating curve), $J_d = 0.8$ (orange squares, orange interpolating curve), 
$J_d = 0.5$ (green rhombi, green interpolating curve), $J_d = 0.2$ (red upward pointing triangles, red interpolating 
curve), and $J_d = 0.1$ (purple downward pointing triangles, purple interpolating curve); \\ 
{\bf b}:   $n_{{\rm st},j}$ computed in an HM with $L=20$, $J=1$,
$\mu_{\rm eff} = 0$, in the large bias limit with $\Gamma_1 = \gamma_L = g = 2$, with the bond impurity symmetrically located between sites $\bar{j} = 10$ and 
$\bar{j}+1=11$,  at  $J_d = 0.5$, and with  (blue dots, blue interpolating curve), $U = 0.5$ (orange squares, orange interpolating curve), 
$U=1.0$ (green rhombi, green interpolating curve), $U=1.5$ (red upward pointing triangles, red interpolating 
curve), and $U=2.0$ (purple downward pointing triangles, purple interpolating curve). }
\label{cmps_2}
\end{figure}

\noindent
We now discuss  the case of a bond impurity, which
we symmetrically locate  in the middle of an even-$L$  chain. Specifically, 
we use the impurity Hamiltonian $H_{\rm bond}$, given by 

\beq
H_{\rm bond} = - \delta J \: \{ c_{\frac{L}{2} }^\dagger c_{ \frac{L}{2} + 1 } + c_{ \frac{L}{2} + 1}^\dagger c_{ \frac{L}{2} } \} 
\:\:\:\: . 
\label{imp.2}
\eneq
\noindent
In Fig.\ref{cmps_2}{\bf a)},   we plot $n_{{\rm st},j}$, computed in an 1HM with $L=20$, $J=1$,
$\mu_{\rm eff} = 0$, and $U=0$, in the large bias limit with $\Gamma_1 = \gamma_L = g = 2$, 
with the bond impurity symmetrically located between sites $\bar{j} = 10$ and 
$\bar{j}+1=11$,  at various values of the total bond strength  $J_d = J + \delta J$ (see figure caption for details). 
At variance, to evidence the effects of the bulk interaction, in    Fig.\ref{cmps_2}{\bf b)} 
we again show  the NESS particle density in real space as a function of the position in the chain, computed for the same parameters as 
in the left-hand plot, except that now we fix $J_d = 0.5$ and vary $U$ from plot to plot 
 (see figure caption for details).

\begin{figure}
\center
\includegraphics*[width=1. \linewidth]{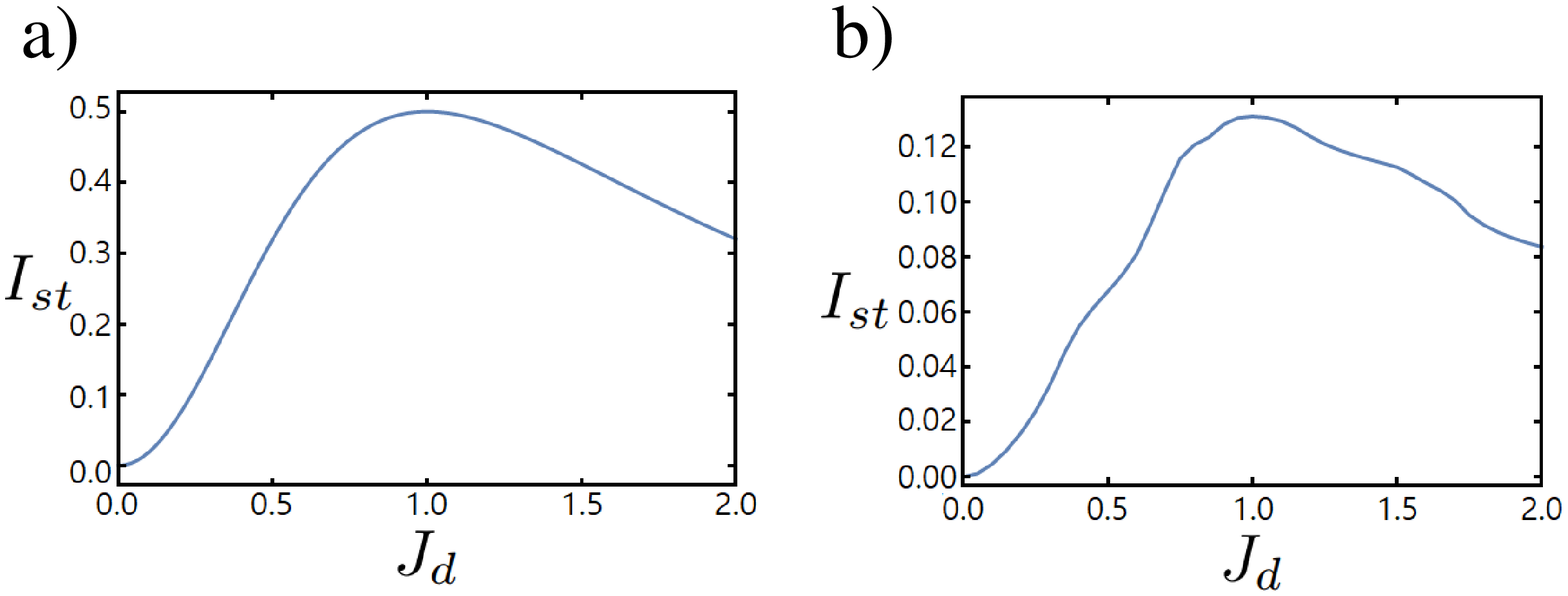}
\caption{{\bf a}: $I_{\rm st}$ as a function of $J_d$ computed in an HM with $L=20$, $J=1$,
$\mu_{\rm eff} = 0$, and $U=0$, in the large bias limit with $\Gamma_1 = \gamma_L = g = 2$,
with the  bond impurity corresponding to the Hamiltonian in Eq.(\ref{imp.2}). 
{\bf b};  The same as in {\bf a)}, but with $U=1.5$. }
\label{curbond}
\end{figure}

\noindent
Aside for quantitative differences, the plots in Fig.\ref{cmps_2} exhibit the same behavior as 
the ones in Fig.\ref{imps_1}. Thus, we conclude that, changing the type of isolated impurity in 
the chain, does not substantially affect the charge-density distribution in the NESS. 
Again, the result in Fig.\ref{cmps_2} is pertinently complemented by looking at  
  $I_{\rm st}$, as a function of $J_d$. Repeating the analysis we have performed above 
in the case of a site impurity, in Fig.\ref{curbond} we plot $I_{\rm st}$ as a function of $J_d$ in the case 
$U=0$ (Fig.\ref{curbond}{\bf a)}), and for $U=1.5$ (Fig.\ref{curbond}{\bf b)}).  Also, in Fig.\ref{cur.1}{\bf b)} we draw plots of 
 $I_{\rm st}$ as a function of $L$ up to $L \sim 100$ and for various values of $\frac{U}{J}$, finding 
 results qualitatively similar to the ones we display in Fig.\ref{cur.1}{\bf a)} for the site impurity.

\noindent
Basically, the same conclusions we reach in the case of a site impurity apply to 
$I_{\rm st}$ in the NESS in the presence of a bond impurity. The main difference,
due to the ``directional'' nature of the bond impurity, compared to 
the site impurity, is that, in this case, the current is not symmetrically 
distributed about $J_d = 1.0$ (corresponding to $\delta J = 0$). 

To summarize, we have provided evidences that a single 
impurity in the chain (either a site impurity, or a bond impurity) does not 
qualitatively affect the NESS with respect to what happens in a homogeneous 1HM. 
So, we expect no relevant modifications in the location and in the 
characteristics of the OWP with respect to the one emerging in the homogeneous chain.

In the next section, we extend this analysis to the case of a finite density of 
impurities in the chain (quenched disorder), particularly focusing on how, and 
to what extent, the emergence of the OWP in the system in the large bias limit 
is affected by the disorder.

\section{NESS and OWP in the presence of a finite density of impurities}
\label{disorder}

In the previous Section  we have argued how  the OWP should not be substantially 
affected by a single localized defect  in the chain. At variance, as it is well estabilished how a finite amount of disorder 
affects the transport properties of the system in the NESS \cite{disorder_1,disorder_2,disorder_3}, we expect that 
it affects the OWP, as well,   in principle even determining it disappearance, 
in the strong disorder limit. Motivated by these observations, in this Section we extend the analysis of the 
effects of the impurities by considering the case in which a finite density of impurities is 
present in the system, by particularly focusing onto the effects of disorder on the OWP. 

To introduce disorder in the 1HM we can, e.g., randomize the chemical potential $\mu$ and/or the bond electron hopping strength $J$ and/or 
the   interaction strength $U$, {\it et cetera}. Yet, apart for differences in the structure of the final phase that are 
realized in the system as a consequence of the disorder (see, for instance, Ref.[\onlinecite{dfisher}] for a comprehensive 
discussion about this point), disorder is, in general, expected to substantially affect the transport properties
of the system, especially in lower dimensions \cite{gang}. Taking this into account and also to be able to make a 
systematic comparison with the results of Ref.[\onlinecite{disorder_1}], in the following we focus onto 
a model with a random chemical potential, corresponding to a random applied field in the $z$-direction in the 
$XXZ$ spin chain discussed in Ref.[\onlinecite{disorder_1}]. Technically, we realize this by 
setting  

\beq
\mu_{\rm eff}  \to \mu_{j}=\bar{\mu} + \delta\mu_{j}
\;\;\;\; , 
\label{dis.1}
\eneq
\noindent
with $j=1,...,L$ and with   $\{ \delta \mu_j \} $ independent 
random variables described by a probability  distribution $ P [ \{ \delta \mu_j \} ] 
= \prod_{ j = 1}^\ell p ( \delta \mu_j )$. Specifically, we choose $p ( \delta \mu )$ to be 
the  probability distribution for $\delta \mu$ with average 
$\bar{\delta \mu} = \int \: d  \delta \mu \: \delta \mu p ( \delta \mu ) = 0$, and with 
variance $\sigma_{\mu}^2 = \int \: d \delta \mu \: \delta \mu^2 p ( \delta \mu ) $. As a result, we obtain

\begin{eqnarray}
 \overline{  \delta \mu_j } &=& \int \: \prod_{ r = 1}^\ell d \delta \mu_r \: P [ \{ \delta \mu_r \} ] \delta \mu_j = 0
 \label{disl.2} \\
  \overline{ \delta \mu_i \delta \mu_j }  &=& \int \: \prod_{ r = 1}^\ell d \delta \mu_r \: P [ \{ \delta \mu_r \} ] \delta \mu_i \delta \mu_j =
 \sigma_{\mu}^2 \delta_{ i , j } \nonumber 
 \:\:\:\: , 
\end{eqnarray}
\noindent
with $\overline{ O [ \{ \delta \mu_j \} ] }$ denoting the ensemble average of a generic functional 
of $\{ \delta \mu_j \}$ with respect to the probability distribution $ P [ \{ \delta \mu_j \} ]$. We  
use the uniform probability distribution given by   

\beq
p ( \delta \mu ) \: = \:  \Biggl\{ \begin{array}{l}
\frac{1}{2 \sqrt{3} \sigma_{\mu}} \;\; , \; {\rm for} \: - \sqrt{3} \sigma_{\mu} \leq V \leq \sqrt{3} \sigma_{\mu} \\
0 \:\: , \: {\rm otherwise}
                 \end{array}
\:\:\:\: . 
\label{disl.4}
\eneq
\noindent
Having 
assumed the probability distribution in Eq.(\ref{disl.4}), we use it to 
estimate the disorder-averaged current distribution at given values of 
$\sigma_\mu$ and $U$. In particular, having stated that, in the ``clean''
limit and at large bias,  the 1HM  is insulating for $\frac{U}{J} > 2$, in the 
following we focus onto the interval of values
$0 \leq \frac{U}{J} \leq 2$. As for what concerns $\sigma_\mu$, we restrict ourselves 
to the interval $0 \leq \frac{\sigma_\mu}{J} \leq 1$ which, as we show in the following, 
for the specific system  we focus on, is enough to trigger a transition to an insulating
phase for any value of $U$, given the values of the system parameters that 
we consider here.

\begin{figure}
\center
\includegraphics*[width=1. \linewidth]{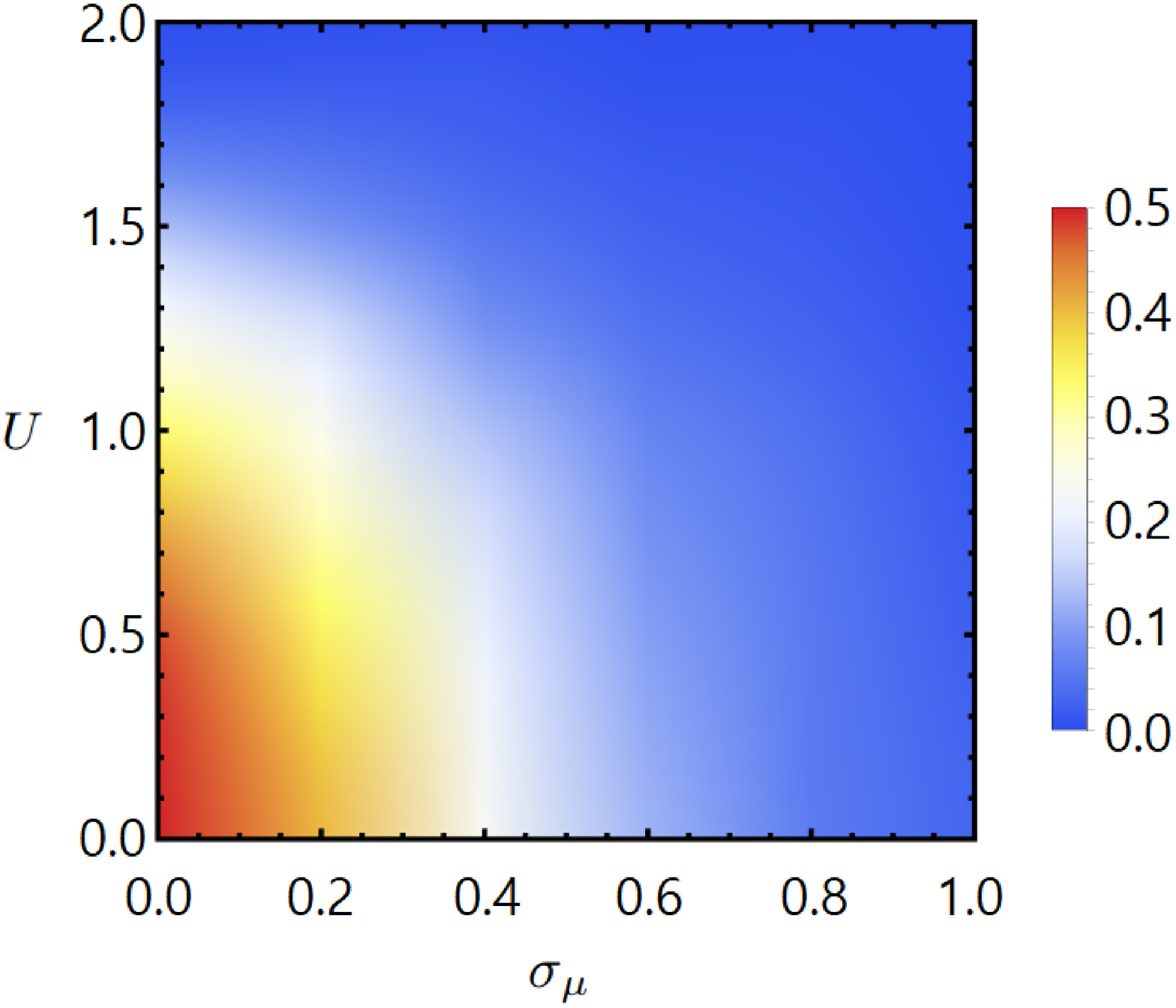}
\caption{$\bar{I}_{\rm st}$ in the 
$\sigma_\mu - U$ plane, obtained by ensemble averaging $I_{\rm st}$ derived in 1HM  with $L=20$-sites with $J=1$, 
$\bar{\mu} = 0$, in the large bias limit with  $\Gamma_1 = \gamma_L = g = 2$,  over $N=50$ realizations of 
the disorder. The color code for the value of $\bar{I}_{\rm st}$ is summarized in the side vertical bar.}
\label{disord_1}
\end{figure}

We report in Fig.\ref{disord_1} our main result for $ \bar{I}_{\rm st}$ in the $\sigma_\mu - U$ plane,  
obtained by ensemble averaging $I_{\rm st}$  
derived in a 1HM  with $L=20$-sites with $J=1$, 
$\bar{\mu} = 0$, in the large bias limit with $\Gamma_1 = \gamma_L = g = 2$,  over $N=50$ realizations of 
the disorder. As a main remark, we note an over-all consistency with the analogous
diagram reported in Fig.1 of Ref.[\onlinecite{noneq}], despite the differences in 
the parameters of the systems considered in the two cases.

\begin{figure}
\center
\includegraphics*[width=1. \linewidth]{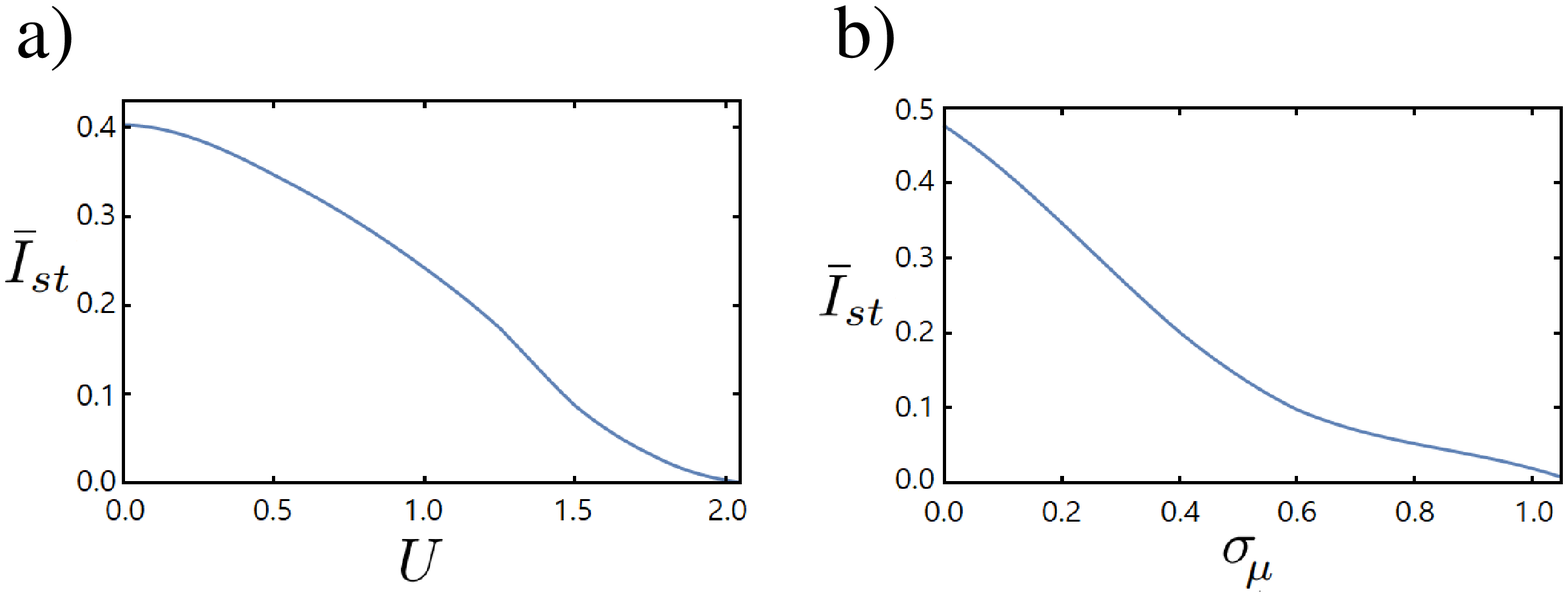}
\caption{\\ {\bf a}: $\bar{I}_{\rm st}$ computed in an $L=20$ chain with $J=1$, $\bar{\mu} = 0$ in 
the large-bias limit with $\Gamma_1 = \gamma_L = g = 2$. along the cut  of Fig.\ref{disord_1} 
corresponding  to the segment $\sigma_\mu = 0.2, 0 \leq U \leq 2$ and to $U=0.5, 0 \leq \sigma_\mu \leq 1$; \\
{\bf b}: $\bar{I}_{\rm st}$ computed in the same system as in {\bf a)} and 
evaluated along  the cut  of Fig.\ref{disord_1} 
corresponding  to the segment  $U=0.5,  0 \leq \sigma_\mu \leq 1$.
}
\label{intdis}
\end{figure}

\noindent
As expected, the larger is $\sigma_\mu$ (at fixed $U$), the lower is the values of $U$ at which the 
transition from the conducting to the insulating phase takes place. This basically gives rise to 
a ``critical line'' in the $\sigma_\mu - U$ plan, with a shading of the transition line due to the 
disorder-triggered nature of the phase transition. Indeed, the conductor to insulator phase transition
can be pictured as a proliferation of the ``kinks'' localized at each defect in the chain, which eventually, 
at a strong enough value of $\sigma_\mu$, coalesce into a single larger kink distributed throughout the 
whole chain. When this happens, 
the conduction gets blocked (similarly to what happens
in the clean system for $\frac{U}{J} > 2$). So, the mechanism appears to be analogous to 
what happens at the Griffiths phase transition in disordered system, with a similar effect 
on the spreading of the critical line between the two phases \cite{gr_1,gr_2,gr_3}. To highlight 
the combined effect of a finite $U$ and a finite $\sigma_\mu$ in triggering the transition to the insulating 
phase, in Fig.\ref{intdis} we plot $\bar{I}_{\rm st}$ along two cuts of Fig.\ref{disord_1}, respectively 
corresponding  to the segment $\sigma_\mu = 0.2, 0 \leq U \leq 2$ (Fig.\ref{intdis}{\bf a)}) and to $U=0.5, 0 \leq \sigma_\mu \leq 1$
(Fig.\ref{intdis}{\bf b)}),
computed in an $L=20$ chain with $J=1$, $\bar{\mu} = 0$, in the large bias limit with  $\Gamma_1 = \gamma_L = g = 2$. 
In both cases, we clearly see the transition from the conducting to the insulating regime.
Apparently, following the analysis of Ref.[\onlinecite{disorder_1}], our result implies that
the localization length of the system, $L_*$, (which is expected to be a function of 
both $\sigma_\mu$ and $U$), is always $\leq L$, so to allow for the disorder-induced transition (in fact a crossover)
to the insulating phase  in the $L=20$ chain.

\begin{figure}
\center
\includegraphics*[width=1. \linewidth]{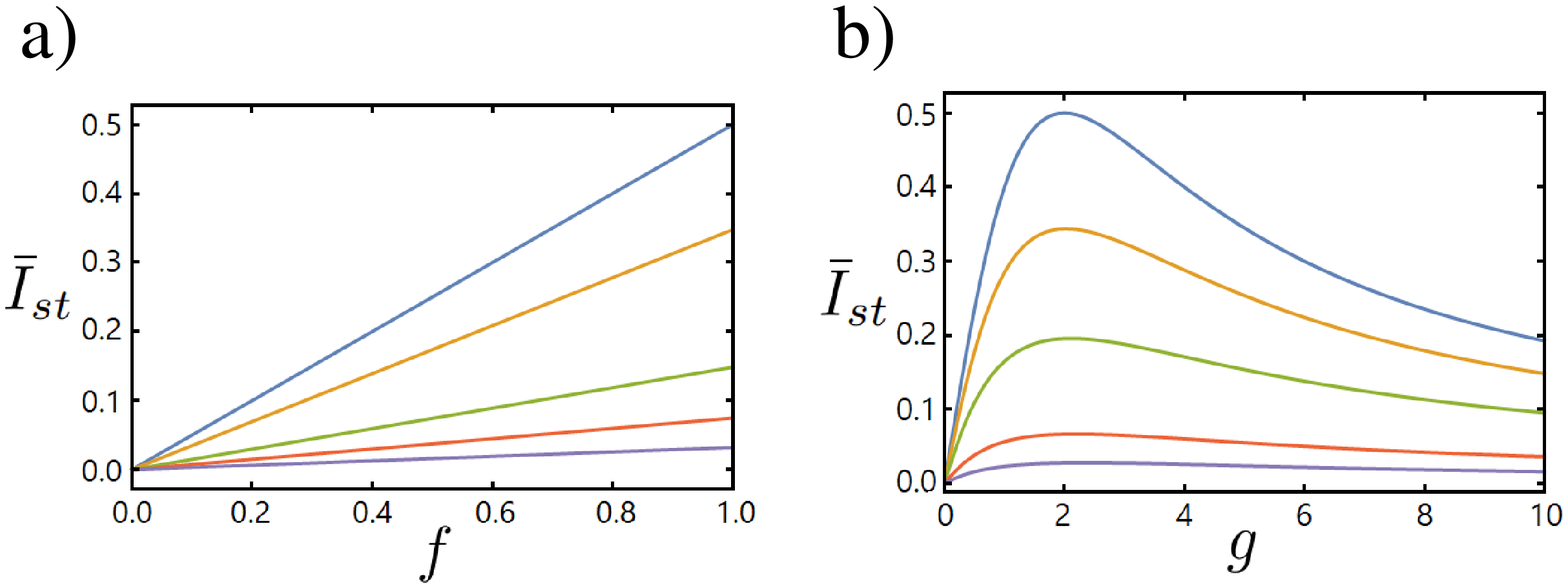}
\caption{\\ {\bf a}: $\bar{I}_{\rm st}$  as a function of $f$  in a 1HM 
 with $L=20$-sites with $J=1$, $\bar{\mu} = 0$, $U=0$,  and $\Gamma_1 = \gamma_L = g = 2$,
computed by ensemble averaging over $N=50$ realizations of 
the disorder for different values of $\sigma_\mu$; \\
{\bf b}: $\bar{I}_{\rm st}$  as a function of $g$  in a 1HM 
 $L=20$-sites with $J=1$, $\bar{\mu} = 0$, $U=0$, taken to the large bias limit 
 with $\Gamma_1 = \gamma_L = g = 2$,
computed by ensemble averaging over $N=50$ realizations of 
the disorder, again for different values of $\sigma_\mu$. In both panels we have set: $\sigma_\mu = 0.0$ (blue line), 
$\sigma_\mu = 0.5$ (yellow line), $\sigma_\mu = 1.0$ (green line), $\sigma_mu = 1.5$ (red line), and 
$\sigma_\mu = 2.0$ (purple line).}
\label{fgu}
\end{figure}

\noindent
Having checked the consistency of our results with the phase diagram of Ref.[\onlinecite{disorder_1}],
we now discuss the effects of the disorder on the OWP. To do so, we first move along the horizontal line 
of Fig.\ref{disord_1} corresponding to $U=0$. Repeating the analysis of Section \ref{chain} in 
the presence of disorder, we compute $\bar{I}_{\rm st}$  as a function of $f$  in a 1HM 
with $L=20$-sites with $J=1$, 
$\bar{\mu} = 0$, $U=0$,   with  $\Gamma_1 = \gamma_L = g = 2$, by ensemble averaging over $N=50$ realizations of 
the disorder and for different values of $\sigma_\mu$ 
(Fig.\ref{fgu}{\bf a)}), as well as at nonequilibrium, as a function of $g$, in the same system and 
using the same procedure, again for different values of $\sigma_\mu$ (Fig.\ref{fgu}{\bf b)}).

\noindent
Remarkably, from Fig.\ref{fgu}{\bf b)}, we see that a limited amount of disorder does not spoil 
our result that the OWP point in $I_{\rm st}$ as a function of $g$ appears in the 
chain when it is taken out of equilibrium  even when $U=0$. Eventually, a strong disorder washes out the
OWP which, from  Fig.\ref{fgu}{\bf a)}, we find to happen simultaneously with 
a reduction of $\bar{I}_{\rm st}$ to 0, that is, to the phase transition from the conducting to the 
insulating phase. 

\begin{figure}
\center
\includegraphics*[width=1. \linewidth]{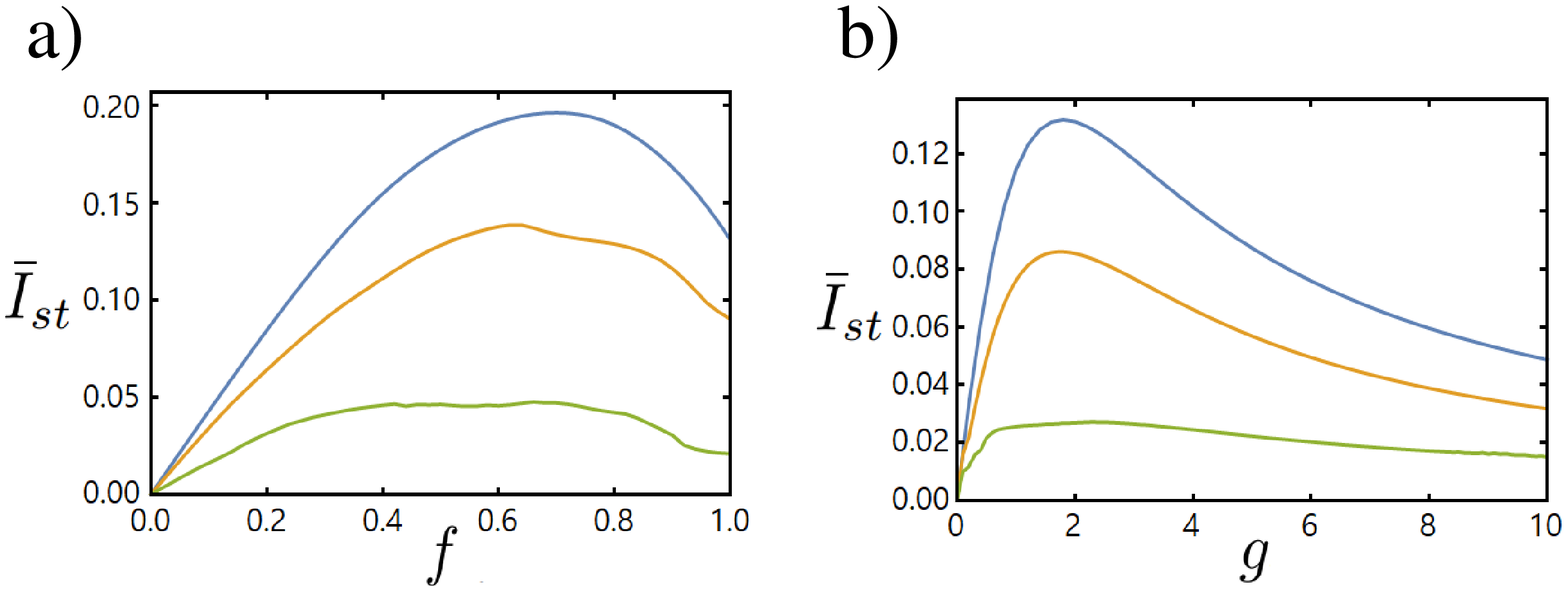}
\caption{\\ {\bf a}: $\bar{I}_{\rm st}$  as a function of $f$  in a 1HM 
at equilibrium with $L=20$-sites with $J=1$, $\bar{\mu} = 0$, $U=1.5$,  and $\Gamma_1 = \gamma_L = g = 2$,
computed by ensemble averaging over $N=50$ realizations of 
the disorder for different values of $\sigma_\mu$; \\
{\bf b}: $\bar{I}_{\rm st}$  as a function of $g$  in a 1HM 
 $L=20$-sites with $J=1$, $\bar{\mu} = 0$, $U=1.5$, taken to the large bias limit 
 with $\Gamma_1 = \gamma_L = g = 2$,
computed by ensemble averaging over $N=50$ realizations of 
the disorder, again for different values of $\sigma_\mu$. In both panels we have set: $\sigma_\mu = 0.0$ (blue line), 
$\sigma_\mu = 0.25$ (yellow line), $\sigma_\mu = 0.5$ (green line).}
\label{fgu1}
\end{figure}

Consistently with our result that a limited amount of disorder does not wash out the OWP at 
$U=0$, we expect that the same happens at $U>0$. To check this guess, in Fig.\ref{fgu1} we 
plot $\bar{I}_{\rm st}$ computed in the same system as the one we used to derive 
Fig.\ref{fgu}, but choosing $U=1.5$. As expected, disorder does not substantially affect 
the OWP, which proves to be pretty stable against the presence of impurities in the chain, both 
in the noninteracting case, as well as for a finite value of $U$.

\noindent
Our sampling analysis, combined with the over-all phase diagram of the disordered 
1HM in the $\sigma_\mu - U$-plane in Fig.\ref{disord_1}, let us infer that, at any point of the phase diagram
characterized by a finite value of the ensemble averaged $\bar{I}_{\rm st}$, it is always possible, in the 
large bias limit, to tune 
the system at the OWP by pertinently operating over the parameters $\Gamma_1 , \gamma_L , g$ and $f$. 
This result, together with our finding that tuning $g$ allows for recovering the OWP   even when $U=0$,
shows that the emergence of the OWP itself and the corresponding onset of a negative differential conductivity 
in the chain, are pretty universal features, robust against both the electronic interaction as well as  the 
disorder in the chain. 
 
\section{Conclusions}
\label{conc}

Using the Lindblad equation approach we have discussed the main features of the NESS arising 
in an interacting one-dimensional electronic chain connected to two reservoirs in the large bias limit. 
To do so, we have characterized the NESS by synoptically monitoring both the stationary
current $I_{\rm st}$ and the stationary charge distribution in real space, $n_{{\rm st}, j}$, 
characterizing the NESS. In the noninteracting case, we were able to do so within a 
fully analytical approach, by providing explicit formulas for $I_{\rm st}$ and 
$n_{{\rm st}, j}$ in the NESS. 

In the presence of a nonzero electronic interaction, we 
have resorted to a   MF approach to the interaction, which allowed us to perform 
a systematic characterization of the NESS as a function of the bias between the reservoirs,
of the strength of the couplings between the chain and the reservoirs, of the bulk interaction
in the chain and, eventually, when breaking the chain homogeneity with an isolated impurity, 
as a function of the type and of the strength of the impurity potential. Finally, 
we have introduced a finite density of impurities in the chain to discuss how the 
NESS depends on the amount of quenched disorder, as well.

Our analysis allowed us to characterize the emergence of an  OWP in the multi-parameter
space, at which $I_{\rm st}$ is maximized with respect to the values of the various 
parameters. Eventually, we showed that the OWP is robust against the presence of a limited
amount of disorder in the chain, while a strong enough disorder washes it out, by 
triggering, at the same time, a disorder-induced transition from a conducting to an 
insulating NESS. 

The importance of our results is strictly related to the importance of 
both understanding the nature of the OWP and, after that, of 
tuning a device at the OWP in a large number of cases of physical interest.

Because of its simplicity, combined with its reliability, which we checked by comparing 
our results to the ones available in the literature about LE approach to nonequilibrium
quantum systems, we plan to extend  our approach  to, e.g., 
look for novel phases/phase transitions arising in  the phase diagram of junctions of interacting fermionic systems \cite{oca,cham,ganpb,kga,nava_0,nava_x1} and/or 
spin chains \cite{tsve_1,gsst,gns_b}, or to define systematical optimization procedure for the parameters 
determining the working point of a quantum device, and so on.

\vspace{0.5cm}

{\bf Acknowledgements --}
We gratefully thank D. Rossini for insightful discussions. 

A. N. was financially supported  by POR Calabria FESR-FSE 2014/2020 - Linea B) Azione 10.5.12,  grant no.~A.5.1.
D. G. and M. R. acknowledge  financial support  from Italy's MIUR  PRIN projects TOP-SPIN (Grant No. PRIN 20177SL7HC).

\appendix

\section{Variational approach to the stationary current and to the real space charge distribution in 
the nonequilibrium stationary state}
\label{varia}

In this appendix we present a simple variational approach to the ``kink confinement'', and 
to the corresponding conductor-to-insulator phase transition, which we found in the 
1HM connected to two reservoirs, in the  large bias limit.  
To do so, we first of all remark that, via the (inverse) Jordan-Wigner transformation, the 
 Hamiltonian in Eq.(\ref{interac.1}) at $\mu_{\rm eff} = 0$ is mapped onto the
Hamiltonian for the  $XXZ$ spin chain in a zero magnetic field, $H_{XXZ}$, given by 

\beq
H_{XXZ} = - J \sum_{ j = 1}^{L-1} \{ S_{j}^+ S_{j+1}^- + S_j^- S_{j+1}^+ \} + 
J \Delta \sum_{ j = 1}^{L-1} S_j^z S_{j+1}^z
\:\:\:\; , 
\label{xxz.1}
\eneq
\noindent
with $S^+_j = c_j^\dagger \: e^{ i \pi \sum_{ t = 1}^{j-1} c_t^\dagger c_t }$ and 
$S_j^z = c_j^\dagger c_j - \frac{1}{2}$. In particular, $\frac{U}{J} < (>) 2$ in the 1HM
corresponds to $\Delta < (>) 1$ in $H_{XXZ}$ in Eq.(\ref{xxz.1}). Along the correspondence between operators 
in the 1HM and in the $XXZ$ Hamiltonian, we find that the charge density at site $j$ and 
the current density through the link between $j$ and $j+1$ in the former model are respectively expressed 
in terms of operators in the latter model as 

\begin{eqnarray}
 n_j &=& S_j^z + \frac{1}{2} \nonumber \\
 I_{j,j+1} &=& - i J \{ S_j^+ S_{j+1}^- - S_{j+1}^+ S_j^- \} 
 \:\:\:\: . 
 \label{xxz.2}
\end{eqnarray}
\noindent
Once the  reservoirs have driven the system towards the NESS, the 
current must be uniform throughout the chain and equal to $I_{\rm st}$. 
At the same time, setting $n_{{\rm st} , j} = \langle n_j \rangle_{\rm st}$, where 
$\langle \ldots \rangle_{\rm st}$ denotes averaging within the NESS, 
in the large bias regime $I_{\rm st}$ is related to 
both $n_{{\rm st}, 1}$ and  $n_{{\rm st}, j}$ through the relations 

\begin{eqnarray}
 I_{\rm st} &=& \Gamma_1 \: [ 1 - n_{{\rm st} , 1}  ] = \Gamma_1 \left\{  \frac{1}{2} - \langle S_1^z \rangle_{\rm st} \right\} \nonumber \\
  I_{\rm st} &=& \gamma_L  \:  n_{{\rm st} , L}  = \gamma_L \left\{  \frac{1}{2} + \langle S_L^z \rangle_{\rm st}  \right\}
  \:\:\:\: , 
  \label{xxz.3}
\end{eqnarray}
\noindent
with the relation with the average local magnetization in the $XXZ$ model explicitly evidenced. 
Choosing, as we have done in Section \ref{interaction}, $\Gamma_1 = \gamma_L$, implies therefore 
$\langle S_1^z \rangle_{\rm st} = -  \langle S_L^z \rangle_{\rm st} \equiv m_{\rm bou}$. 
Resorting to the $XXZ$ spin chain framework we can therefore adapt the semiclassical approach 
introduced in Ref.[\onlinecite{alcaraz}] to discuss the kink dynamics for $\Delta \geq 1$ to 
a generic value of $\Delta$ and, in particular, to the case $| \Delta | \leq 1$, which is the 
one we focus on in  Section \ref{interaction}. 

Following  Ref.[\onlinecite{alcaraz}], we treat the spin operators in $H_{XXZ}$ as classical 
variables, for which we make an appropriate variational ansatz. Computing the energy of 
the corresponding state using Eq.(\ref{xxz.1}) supplemented with the constraints implied 
by Eqs.(\ref{xxz.3}) and minimizing the corresponding result with respect to the variational 
parameters, we find out how the kink solution varies as a function of the $I_{\rm st}$ as well 
as of  $U$. Letting $\vec{\cal S}_j$ be our variational ansatz for the spin operator
at site $j$ (to be eventually identified with $\langle \vec{S}_j \rangle_{\rm st}$), we set 
(assuming that, at the boundaries, the spin are fully polarized in the $z$-direction, due to 
the coupling with the reservoirs)

\beq
  \vec{\cal S}_j \equiv  \left[ \begin{array}{c}
{\cal S}_j^x \\      {\cal S}_j^y \\       {\cal S}_j^z               
                        \end{array} \right]  
                          = 
\left[ \begin{array}{c}
\cos ( \varphi_j ) \sin ( \vartheta_j ) \\ 
\sin ( \varphi_j ) \sin ( \vartheta_j ) \\ 
 \cos ( \vartheta_j )
       \end{array}
   \right]
\:\:\:\: , 
\label{xxz.4}
\eneq
\noindent
with $\vartheta_j , \varphi_j$ being smooth functions of the real space variable, so to 
enable us to treat them as smooth functions of a continuous coordinate variable $x$.  
In order to choose ${\cal S}_j^z = \cos ( \vartheta_j )$ so to match the density 
profiles in Fig.\ref{rea_spa_den}, we make a ``minimal''  variational ansatz, 
that is, we fit the magnetization profile  with a trial function depending on one variational 
parameter only. As we discuss below, though being a pretty crude approximation, our 
variational ansatz is apparently good enough to allow for qualitatively recovering 
all the key features highlighted in Sections \ref{chain},\ref{interaction}. Specifically, we choose our  trial function 
so that (resorting to a the continuous variable $x$)

\begin{eqnarray}
{\cal S}_j^z &\to&     \frac{1}{2}   \: \left\{  \frac{1}{1 + e^{  (x-\ell) } }
- \frac{1}{1 + e^{  (L-x-   \ell)  } }\right\}   \nonumber \\
&\equiv & \frac{ 1 }{2}    \: \cos ( \vartheta ( x = a j  ) ) 
\:\:\:\: , 
\label{xxz.5}
\end{eqnarray}
\noindent
with $a$ being the lattice step and $L$ being the length of the chain. The only variational parameter entering the function in Eq.(\ref{xxz.5})
is the length scale $\ell$.   

\begin{figure}
\center
\includegraphics*[width=1 \linewidth]{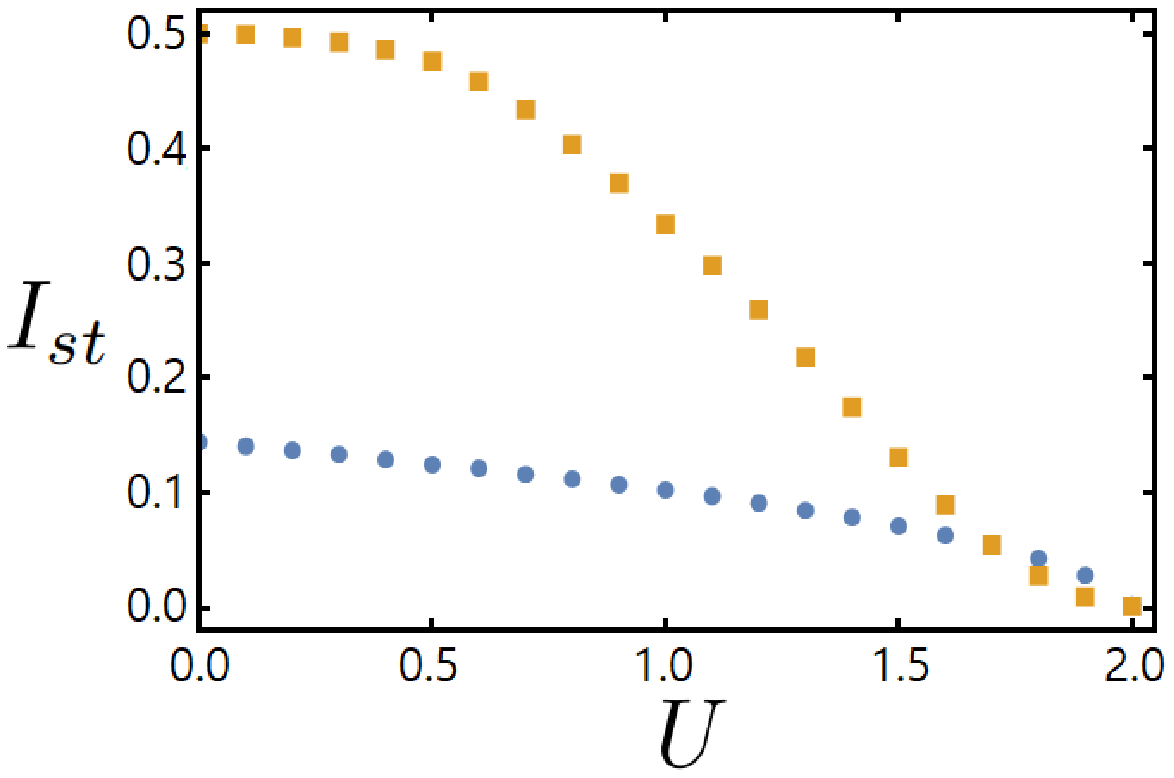}
\caption{ $I_{\rm st}$  as a function of 
$U$  computed within our variational approach in an $L=20$  chain with $J=1$ and  $\mu_{\rm eff} = 0$, in
the large bias limit with   $\Gamma_1 = \gamma_L = g = 2$ (blue dots),  and 
using the numerical approach of Section \ref{interaction} with the same values of the system 
parameters (orange squares). }
\label{varcur}
\end{figure}

To variationally determine $\ell$, we  estimate the energy for the Hamiltonian in Eq.(\ref{xxz.1}) corresponding to our variational solution in Eq.(\ref{xxz.5}), ${\cal E} [ \ell ]$,  
by using the MF result of Ref.[\onlinecite{saleur}]. Doing so, we set 

\begin{eqnarray}
 && {\cal E} [ \ell ]  \approx   \frac{J}{2}   \int_0^L   d x \left\{ \left( \frac{d \vec{\cal S}}{d x } \right)^2 - 2 ( \Delta - 1) ( {\cal S}^z)^2 \right\} \nonumber \\
 &=& \frac{J}{2}   \int_0^L   d x \biggl\{ \left( \frac{d \vartheta(x) }{d x } \right)^2  + \sin^2 ( \vartheta ( x )) \left( \frac{d \varphi ( x ) }{d x } \right)^2 \nonumber \\
 &-& 2 ( \Delta - 1) \cos^2 ( \vartheta ( x ))  \biggr\} 
 \:\:\:\: . 
 \label{xxz.8}
\end{eqnarray}
\noindent
In addition to $\vartheta ( x )$,  ${\cal E} [ \ell ]$ also depends on  $\varphi ( x )$. To determine this latter function
we employ the current conservation within the NESS, which implies that $I_{\rm st}$ is the same 
independently of the position in the chain. Within MF approximation  \cite{saleur}, we get 
 $I_{\rm st} = J \: \sin^2 ( \vartheta ( x ) ) \frac{d \varphi ( x ) }{d x }$.
Therefore, we obtain

\begin{eqnarray}
I_{\rm st} &\approx&  J \: \sin^2 ( \vartheta ( x ) ) \frac{d \varphi ( x ) }{d x } \nonumber \\
&\Rightarrow&  \frac{d \varphi ( x ) }{d x } = \frac{I_{\rm st}}{J  } \: \sin^{-2} ( \vartheta ( x )) 
\:\:\:\: .
\label{xxz.6}
\end{eqnarray}
\noindent
Using Eq.(\ref{xxz.6}), we may rewrite Eq.(\ref{xxz.8}) as

\begin{eqnarray}
  {\cal E} [ \ell ]  
 &=& \frac{J}{2}   \int_0^L   d x \biggl\{ \left( \frac{d \vartheta(x) }{d x } \right)^2  + \left( \frac{ I_{\rm st} }{J  \sin^2 ( \vartheta ( x ))  } \right)^2 \nonumber \\
 &-& 2 ( \Delta - 1) \cos^2 ( \vartheta ( x ))  \biggr\} 
 \:\:\:\: . 
 \label{xxz.9}
\end{eqnarray}
\noindent
To explicitly put the right-hand side of Eq.(\ref{xxz.9}) in a form depending on 
$\ell$ only, we recall  that, from Eqs.(\ref{xxz.3}), we obtain    

\beq
I_{\rm st} = \Gamma_1 \: \left\{ \frac{1}{2} - m_{\rm bou} \right\} 
=  \Gamma_1 \: \left\{ \frac{1}{2} - \frac{ \cos ( \vartheta ( 0 )) }{2}  \right\} 
\:\:\:\: . 
\label{xxz.7}
\eneq
\noindent
To recover the results of Figs.\ref{iint},\ref{rea_spa_den}, we minimized, with respect to $\ell$, ${\cal E} [ \ell ]  $
computed at fixed $\Gamma_1 = \gamma_L = g = 2$ and for a given $\Delta$. Once we had estimated in 
this way the parameter in the  trial function of Eq.(\ref{xxz.5}), we computed  $n_{{\rm st} }$ throughout the chain as a function of 
$U$, as well as $m_{\rm bou}$. Knowing $m_{\rm bou}$, we eventually used Eq.(\ref{xxz.7}) to compute $I_{\rm st}$  as a function of 
$U$. In Fig.\ref{varcur} we draw $I_{\rm sp}$ computed within our variational approach   as a function of $U$ in 
the large bias limit with  $\Gamma_1 = \gamma_L = g = 2$ (blue dots), together with the analogous quantity computed using the numerical 
  approach of Section \ref{interaction} (orange squares), in an $L=20$  chain with $J=1$ and  $\mu_{\rm eff} = 0$. 
  While, possibly due to our oversimplified choice for the 
trial wavefunction, there is a rather weak quantitative agreement between the points, 
we believe that the qualitative agreement is satisfactory enough and witnesses the reliability of our method: indeed, we 
see that  in both cases   
$I_{\rm st}$ monotonically decreases on increasing $U$ from 0 to positive values and eventually becomes 0 as soon as 
$U \geq 2$, which is also consistent with Refs.[\onlinecite{rossini_0,rossini}]. Moreover, as pointed out in 
Ref.[\onlinecite{saleur}], the effective continuum energy functional in Eq.(\ref{xxz.8}) is expected to be mostly reliable 
for $\Delta \sim 1$, corresponding to $\frac{U}{J} \sim 2$ in our model, where, indeed, the agreement between the two 
plots is pretty good, even quantitatively.

\begin{figure}
\center
\includegraphics*[width=1 \linewidth]{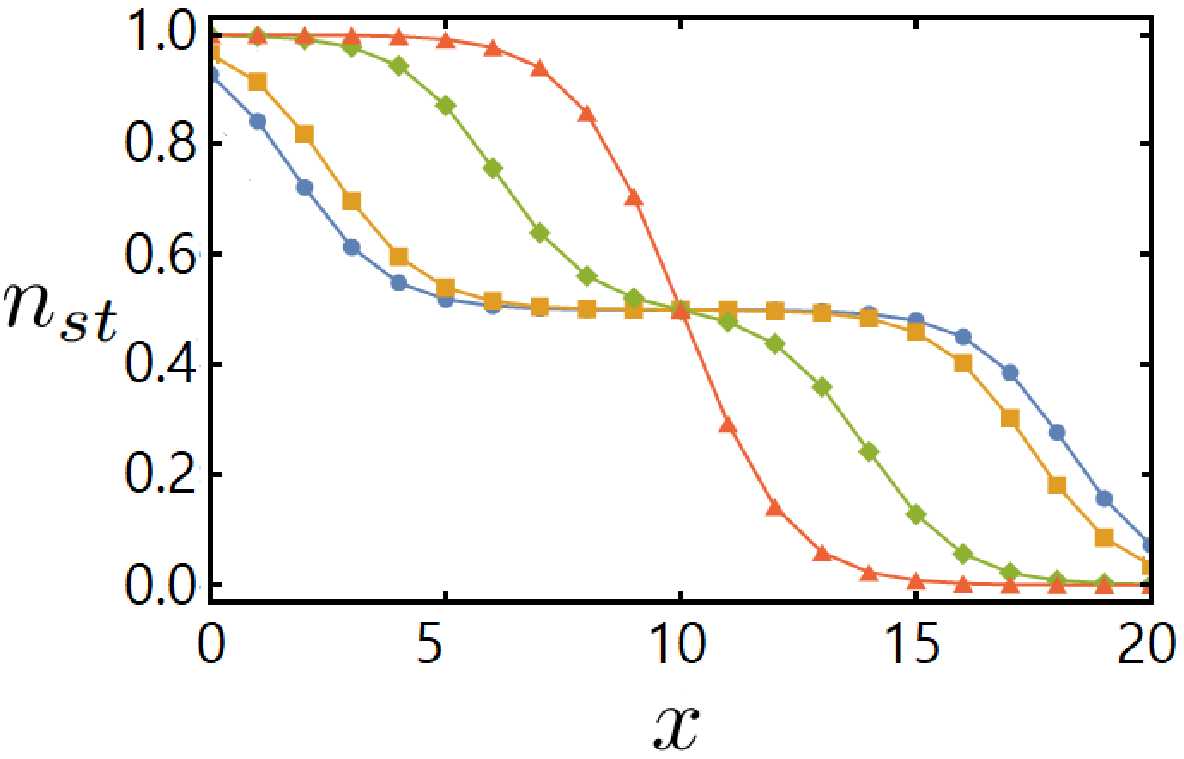}
\caption{ $n_{\rm st}$  as a function of $x$ 
computed within our variational approach in an $L=20$  chain with $J=1$ and  $\mu_{\rm eff} = 0$, in
the large bias limit with   $\Gamma_1 = \gamma_L = g = 2$, and with 
$U=0$ (blue circles - blue interpolating curve),
$U=1.5$ (orange rhombi - orange interpolating curve),
$U=1.9$  (green squares - green interpolating curve), and  
 $U=2.1$ (red triangles - red interpolating curve). 
 Note that, to account for errors induced by the crude approximations in our choice of the trial function, 
 the values of $U$ are slightly larger than what is expected from the numerical results of 
 Section \ref{interaction}. Yet, the qualitative agreement with the numerical results is pretty good.}
\label{dens_var} 
\end{figure}

\noindent
In Fig.\ref{dens_var}, we plot $n_{\rm st}  $ within the NESS as a function of $x$ computed within the variational approach. Apparently, 
using the variational approach allows us for recovering 
a pretty good agreement with the trend evidenced in Fig.\ref{rea_spa_den}: as long as the system supports a nonzero $I_{\rm st}$, 
$n_{\rm st}$ is flat at $n_{\rm st} = \frac{1}{2}$ throughout the middle part of the chain, with a respectively upward and downward turn
close to the endpoints of the chain, that are required to match $n_{{\rm st} , 1}$ and $n_{{\rm st} , L}$ as determined by the constancy 
of $I_{\rm st}$. On increasing $U$, the extent of the flat region gets reduced, till the region shrinks at $U = 2$, where 
$n_{\rm st}$ takes a ``kink-like'' profile, with a corresponding blocking of the current transport ($I_{\rm st} = 0$), for 
$U > 2 $ \cite{rossini,rossini_0}.

\noindent
 
\begin{figure}
\center
\includegraphics*[width=1 \linewidth]{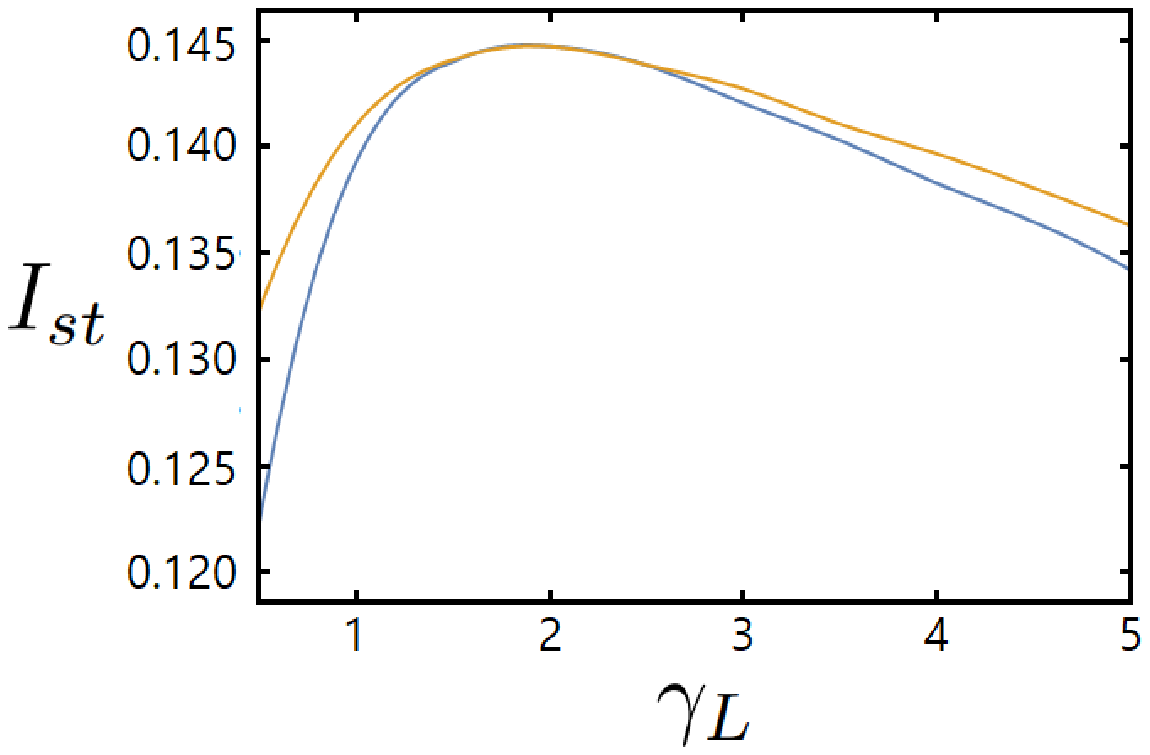}
\caption{$I_{\rm st}$ computed using the variational approach at $J=1$, $L=20$ and $U=0$, at large bias,
as a function of $\gamma_L = \Gamma_1 = g$ (blue curve) and as a function of $\gamma_L = g$ at $\Gamma_1 = 2$
(orange curve). The reduction by a factor of 4 with respect to the numerical results of 
Section \ref{chain} is consistent 
with the analogous reduction of $I_{\rm st}$ {\it vs.} $U$ in Fig.\ref{varcur} as $U \to 0$.}
\label{owp_var} 
\end{figure}

Before concluding this Appendix, it is worth mentioning that, by means of a simple extension of the 
crude variational approach we discussed above, we are able to catch the remarkable emergence of an 
OWP in the plot of $I_{\rm st}$ as a function of $\Gamma_1 = \gamma_L = g$, derived in the fully nonequilibrium 
limit $f=1$ and in the noninteracting case $U=0$. In fact, while, consistently with 
Refs.[\onlinecite{rossini}], in Section \ref{interaction} we find  no OWP when plotting $I_{\rm st}$ as a function of 
$f$ for $\Gamma_1 = \gamma_L = g$ and for $U=0$, instead, we do find the OWP in the plot of
$I_{\rm st}$ as a function of $\Gamma_1 = \gamma_L = g$, as we show throughout Section \ref{chain}, as well as 
in  Fig.\ref{fig:interaction_plot}{\bf b)}. To recover the OWP for $U=0$ we used   the energy 
functional in Eq.(\ref{xxz.8}) with $\Delta = 0$ and a variational function  obtained by applying
a rigid translation by $x_0$ to the function of Eq.(\ref{xxz.5}), that is 

\begin{eqnarray}
{\cal S}_j^z &\to&     \frac{1}{2}   \: \left\{  \frac{1}{1 + e^{  (x-x_0-\ell) } }
- \frac{1}{1 + e^{  (L-x + x_0-   \ell)  } }\right\}   \nonumber \\
&\equiv & \frac{ 1 }{2}    \: \cos ( \vartheta ( x = a j  ) ) 
\:\:\:\: .
\label{xxz.a1}
\end{eqnarray}
\noindent
We determine the parameters $\ell$ and $x_0$  by imposing the constraint that 
the values of $m_{{\rm bou},1} , m_{{\rm bou},L}$ obtained from Eq.(\ref{xxz.a1}) are 
consistent with Eq.(\ref{xxz.3}). Doing so, we obtain the plot in Fig.\ref{owp_var}, where 
we draw $I_{\rm st}$ computed using the variational approach at $J=1$, $L=20$ and $U=0$, at large bias,
as a function of $\gamma_L = \Gamma_1 = g$ (blue curve) and as a function of $\gamma_L = g$ at $\Gamma_1 = 2$
(orange curve). In both cases the location of the OWP is consistent with the results of Section \ref{chain}, 
though, due to the pretty crude approximations behind our variational approach, the calculated value of 
$I_{\rm st}$ is lower than the one numerically computed by a factor of 4. Yet, this is consistent 
with the analogous reduction of $I_{\rm st}$ {\it vs.} $U$ in Fig.\ref{varcur} as $U \to 0$.

\noindent

\FloatBarrier
\bibliography{paper_biblio.bib}
\end{document}